\documentclass[preprint]{aastex}
\usepackage{graphicx}

\begin{document}

\title{A Census of X-ray gas in NGC 1068:  Results from 450ks of $Chandra$ HETG 
Observations}

\author{T. Kallman}

\affil{NASA/GSFC}
	
\author{ Daniel A. Evans}

\affil{Harvard-Smithsonian CfA }

\author{ H. Marshall, C.  Canizares, M. Nowak, N. Schulz}

\affil{MIT}

\author{ A. Longinotti}

\affil{European Space Astronomy Center of ESA, Madrid, Spain}

\begin{abstract}
We present models for the X-ray spectrum of the Seyfert 2 galaxy 
NGC 1068.  These are fitted to data obtained using the 
High Energy Transmission Grating (HETG) on the $Chandra$ X-ray observatory.
The data show line and radiative recombination continuum (RRC) 
emission from a broad range of ions and elements.  The models 
explore the importance of excitation processes for these lines 
including photoionization followed by recombination, 
radiative excitation by absorption of continuum radiation 
and inner shell fluorescence.  The models show that the 
relative importance of these processes depends on the 
conditions in the emitting gas, and that no single 
emitting component can fit the entire spectrum.  In particular, 
the relative importance of radiative excitation and 
photoionization/recombination differs according to the 
element and ion stage emitting the line.  This in turn 
implies a diversity of values for the ionization parameter of the 
various components of gas responsible for the emission, 
ranging from log($\xi$)=1 -- 3.  
Using this, we obtain an estimate for the total amount of 
gas responsible for the observed emission.   The mass flux through the 
region included in the HETG extraction region is approximately 
0.3 M$_\odot$ yr$^{-1}$ assuming ordered flow at the speed characterizing 
the line widths.  This can be compared 
with what is known about this object from other techniques.
\end{abstract}

\section{Introduction}

X-ray spectra demonstrate that many compact sources are viewed
through partially ionized gas.  This gas manifests itself 
as a rich array of lines and bound-free features in the 0.1 - 10 keV 
energy range.  The existence of this `warm absorber' 
gas and the fact that it often shows Doppler shifts indicating outflows, 
has potential implications for the mass and energy budgets of these 
sources.   These features have been used to infer mass outflow rates for 
many accreting sources, notably those in bright, nearby active galactic 
nuclei (AGN) \citep{Turn09,Cren03,Mill09,Kron03, Kaas11, Kasp02}.

Intrinsic to the problem is the fact that the gas 
is not spherically distributed around the compact object.  If it were,
resonance lines would be absent or would have a P-Cygni character.
P-Cygni profiles are observed in X-ray resonance lines from the 
X-ray binary Cir X-1 \citep{Schu08}; most AGN do not show this 
behavior clearly.  Understanding the quantity of warm absorber gas, its origin and fate,
are key challenges.  
Interpretation of absorption spectra is often based on 
the assumption that the residual flux in the line trough is solely 
due to finite optical depth or spectral  resolution.  
That is, the effect of filling in by emission is neglected.  
Understanding the possible influence of emission is desirable, 
both from this point of view and since emission provides 
complementary information about a more extended region.
Comparison  between warm absorbers and the spectra of objects 
in which the gas is viewed in reflection rather than 
transmission provides an added test for our 
understanding of the geometry of the absorbing/emitting gas.
Emission spectra also are possibly less affected by systematic 
errors associated with the observation, such as internal background 
or calibration errors which could affect the residual flux in the 
core of deep absorption features.  

Emission spectra are observed from compact objects when the direct 
line of sight to the central compact object is blocked.
This can occur in X-ray binaries or in Seyfert galaxies 
where the obscuration comes from an opaque torus \citep{Anto85}.  A notable 
example is in Seyfert 2 galaxies, and the brightest such object 
is NGC 1068 \citep{Blan97}.  This object has been observed by every X-ray 
observatory with sufficient sensitivity, most recently by 
the grating instruments on $Chandra$ and $XMM-Newton$.   These reveal 
a rich emission line spectrum, including lines from highly ionized 
medium-Z elements, fluorescence from near neutral material, and 
radiative recombination continua \citep{Lied90} (RRCs) which are
indicative of recombination following photoionization.  

Apparent emission can be caused by various physical mechanisms.
These include cascades following recombination, electron impact excitation,
inner shell fluorescence, and radiative excitation by absorption 
of the continuum from the central object in resonance lines.  
Radiative excitation
produces apparent emission as a direct 
complement to the line features seen in warm absorbers; 
rates can exceed that due to other processes.
It is also often referred to as resonance scattering,
since it is associated with scattering of continuum photons 
in resonance lines.  However, past treatments in the 
context of Seyfert galaxy X-ray spectra have been limited
to consideration of a single scattering of an incident continuum 
photon by a resonance line, after which the photon is assumed to be 
lost.  Since there is a considerable literature on the 
topic of transfer of photons in resonance lines, also 
sometimes referred to as resonance scattering, here and in what 
follows we will not use this term.

Since radiative excitation preferentially affects lines
with large oscillator strengths arising 
from the ground term of the parent ion, it affects line ratios
such as the $n=$1 -- 2 He-like lines, causing them to resemble
the ratios from coronal plasmas.  
This was pointed out by \citet{Kink02}, who presented 
high resolution spectra of NGC 1068 obtained using the 
reflection grating spectrometer (RGS) on the 
$XMM-Newton$ satellite.  The RGS is most sensitive at 
wavelengths greater than approximately 10 $\AA$, 
and \citet{Kink02} showed that the He-like lines 
from N and O in NGC 1068 are 
affected by radiative excitation.   They also showed 
that, since radiative excitation depends on pumping by 
continuum radiation from the central compact source, this process 
is affected by the attenuation of the continuum.  
This in turn depends on the column density of the 
line emitting gas;  radiative excitation is suppressed 
when the continuum must traverse a high column density.
\citet{Kink02} adopted a simple picture in which the 
gas is assumed to be of uniform ionization and opacity
and were able to then constrain the column density in 
the NGC 1068 line emitting region from the observed line ratios.

The High Energy Transmission Grating (HETG) on the $Chandra$ 
satellite is sensitive to wavelengths between 1.6 -- 30 $\AA$, 
allowing study of the He-like ratios from elements Ne, Mg, Si 
and heavier, in addition to those from O.   \citet{Ogle03} 
used an HETG observation of NGC 1068 to show that the 
He-like ratios from Mg and Si show a stronger signature of 
radiative excitation than do O and Ne.  They interpret this 
as being due to the fact that all the lines 
come from the same emitting region with a high column, and 
that attenuation of the continuum is stronger for Mg and Si owing 
to ionization effects.  That is, that O and Ne are more highly 
ionized than the heavier elements, and point out that 
this is consistent with the results of photoionization models.

The challenge of modeling X-ray emission spectra divides into 
parts:  (i) modeling the excitation process which leads to emission; 
(ii) modeling the ionization of the gas; and (iii) summing 
over the spatial extent of the emitting gas  leading to the observed 
spectrum.   \citet{Kink02}
include a detailed treatment of the population kinetics for 
emitting ions which takes into account the effect of the 
continuum radiation from the compact object on the populations, 
along with the attenuation of this radiation in a spatially extended 
region.  \citet{Ogle03} add to this a treatment of the ionization 
balance.
\citet{Matt04} have used the intensities of lines formed by fluorescence to 
infer the covering faction of low ionization, high column density gas
and shown that this is consistent with the properties of the obscuring 
torus.

A shortcoming of efforts so far is that none self-consistently 
treats the spatial dependence of the 
absorption effects affecting radiative excitation 
together with the corresponding effects on the ionization balance.
That is, models such as \citet{Ogle03}  adopt temperature and ionization 
balance associated with optically thin photoionized gas.  They then 
assume that this ionization is constant throughout the cloud 
and use this to calculate continuum attenuation and its 
effects on the population kinetics.  They did not 
consider the spatial dependence of the ionization of the gas associated with 
attenuation of the incident continuum, along with the associated 
suppression of radiative excitation.

Since the work of \citet{Kink02} and \citet{Ogle03} there have been numerous 
studies of absorption dominated warm absorber spectra.  In addition, 
a large campaign with the $Chandra$ HETG was carried out on NGC 1068 
in 2008, resulting in a dataset with unprecedented statistical accuracy
\citep{Evan10}.  For these reasons, and the reasons given above, 
we have carried out new modeling of the photoionized emission spectrum 
of NGC 1068.  The questions we consider include: To what extent can 
photoionized emission models fit the observations?  
Which structures, known from other studies, can account 
for the observed X-ray line
emission?  Are there patterns in the elemental abundances in the
photoionized emission spectrum which provide hints about the origin 
or fate of the X-ray gas?  What mass of gas is associated with the X-ray 
emission, and what flow rate does this imply?
In section \ref{data} we present the observed spectrum and 
in section \ref{model} we describe model fitting.  
A discussion is in section \ref{discussion}.

\section{Data}
\label{data}

The dataset that we use in this paper consists of 
an approximately 400 ksec observing campaign on NGC 1068 
during 2008 using the $Chandra$ HETG.  A description of these observations 
was reported 
briefly in \citet{Evan10}, but no detailed description of the data 
has been published until now.  
In addition, we also incorporate
the  earlier 46 ksec observation with the same instrument which was 
carried out in 2000 and was published by \citet{Ogle03}.  
A log of the observations is given in table \ref{table1}.

The satellite roll angles during the observations  were in the range 
308 -- 323 degrees.  This corresponds
to the dispersion direction being approximately perpendicular to the 
extended emission seen in both optical and X-rays \citep{Youn01}.
The standard HEG and MEG extraction 
region size is 4.8 arcsec on the sky.  This is comparable to the 
extent of the brightest part of the X-ray image, which is approximately 
6 arcsec.  It is possible to extract the spectra from regions which 
are narrower or wider 
in the direction perpendicular to the dispersion, and we will 
discuss this below.

The spectrum was extracted using the standard Ciao tools, including 
detection of zero order, assigning grating orders, 
applying standard grade filters, gti filters, and making response files.
The dispersion axes for all the observations were nearly 
perpendicular to the axis of the narrow line region 
(NLR; position angle $\simeq$ 40 degrees). 
As pointed out by \citet{Ogle03} the width of the nuclear emission 
region in the dispersion direction for the zeroth order image is 0.81 -- 0.66 
arcsec, which corresponds to a smearing of FWHM = 0.015-0.018 
Å over the 6-22 Å range in addition to the instrumental profile ( FWHM = 
0.01, 0.02 Å for HEG, MEG). Events were filtered by grade according to 
standard filters, streak events were removed. First-order HEG 
and MEG spectra of the entire dataset were extracted using CIAO 4.4. 
Positive and negative grating orders were added.

In this paper we analyze data extracted using the standard 
size extraction region from the HETG, which is 1.3 $\times 10^{-3}$ degrees or 
4.8 arcsec.  Significant line emission does originate from the 
`NE cloud', located 3.2 arcsec from the nuclear region,
and was discussed by \cite{Ogle03}.  This emission is a factor 2-3 
times weaker than the emission centered on the nucleus.
In section \ref{spatialsec} we briefly discuss analysis of 
spatially resolved data, i.e. data from smaller or larger 
extraction regions.  A more complete examination of the spatial 
dependence of the spectrum will be carried out in a subsequent paper.

We fit the spectrum with a model consisting of 
an absorbed power law continuum plus Gaussian lines.  Table \ref{table2}
lists 86 lines, of which 67 are distinct features detected using the criteria
described below.  
Gaussian line fitting was carried out using an automated procedure
which fits the spectrum in 2$\AA$ intervals.  These are chosen for convenience:
narrow enough that the continuum can be approximated by power law and 
containing a small number of features.  Adjacent intervals 
are chosen to overlap in order to avoid artifacts associated with boundaries.  
Within each interval we fit a local power law continuum plus Gaussians.  
The Gaussians are added to the model at wavelengths corresponding to known 
lines.  For each chosen centroid wavelength the width and normalization 
are varied using a $\chi^2$ minimization procedure until a best fit is 
obtained.  The centroid wavelength is also allowed to vary by a moderate 
amount, twice the value of a fiducial thermal Doppler width, 
during this procedure.  
The trial line is considered to be detected if the $\chi^2$ 
improves by 10, corresponding 
to approximately 99$\%$ confidence for 3 interesting parameters \citep{Avni76}.  
The fiducial thermal Doppler width is a free parameter in this procedure 
(but note it does not influence the final fitted width, only 
the searching procedure for the line center).  
For the results shown here we adopt a value of 200 km s$^{-1}$ for this 
parameter.  Most lines are free of blending 
so the results of the line search are independent of the value of 
this parameter.

Also included in table \ref{table2} are tentative identifications 
for the lines, along with laboratory wavelengths.  These come from the 
{\sc xstar} database \citep{Kal01,Baut01}.  We treat as potential  
identifications the (likely) strongest feature which lies within a wavelength range
corresponding to a Doppler shift of $\leq$+5000 km s$^{-1}$ 
from a known resonance line or RRC.  Table \ref{table2} includes
the values of the implied Doppler shift for these identifications.
Note that our detection criterion does not guarantee accurate measurement 
of the other line parameters, i.e. centroid wavelength and width, so that
weaker lines do not all have bounds on these quantities.  Where they 
are given in table \ref{table2}, errors on these quantities and the line 
flux represent 90$\%$ confidence ($\Delta\chi^2\simeq$3 for one parameter; 
\citet{Avni76}) limits.

The  NGC 1068 spectrum together with the model fits discussed in the next 
section are shown in 
Figures \ref{fig1a} -- \ref{fig1d}.   These are plotted in 2 $\AA$ intervals
and separated according to grating arm (HEG or MEG).  We do not display 
wavelength regions where the grating arm has no sensitivity, or where 
there are no features.  No rebinning or grouping is performed 
in these plots or in our fitting procedure.  We do quote values of 
$\chi^2$ in section \ref{model}, and these were calculated using 
binned data.  Also,  when 
plotting the longer wavelength regions, we bin the spectrum 
is binned for plotting purposes only in order to avoid the 
dominance of the noisy bins with low counts at long wavelengths 
(i.e. $\geq$18 $\AA$).
 In addition, the iron K region, 1.5 -- 2.5 $\AA$ is 
not plotted here, and is discussed in section \ref{feksection}.

Notable line features in the spectrum include the 1s-np lines of H- and He-like ions of
O, Ne, Mg, Si and S.  Corresponding features are also present from Ar and 
Ca, though not clearly detected.  
The 21 -- 23 $\AA$ wavelength region contains the He-like lines 
of oxygen.  These are discussed in more detail in section \ref{helines} below.
The 17 -- 19 $\AA$ wavelength region contains the L$\alpha$ line from
H-like line of O VIII, the most prominent line in the spectrum.  In addition, 
lines from Fe XVII near 16.8 and 17.1 $\AA$, and n=1 -- 3, 1--4 and 1 --5 lines
from O VII near 17.4, 17.8 and 18.6 $\AA$, respectively.  
These lines are discussed  in section \ref{fellines}.
The 15 -- 17 $\AA$ wavelength region contains the L$\beta$ lines from O VIII
plus lines from ion stages of iron:  Fe XVII -- XIX. The n=1 -- 2 lines from 
He-like Ne are at 13.5 -- 13.8 $\AA$.
The 11 -- 13 $\AA$ wavelength region contains the 
L$\alpha$ lines from Ne IX at 12.16 $\AA$, 
the higher order n=1 -- 3 and n=1 --4 lines from Ne IX at 11.6 and 11.0 $\AA$, 
respectively, and additional lines from Fe XVII -- XIX.  Also apparent 
between 11.15 and 11.35 $\AA$ are inner shell fluorescence lines from 
L shell ions of Mg:  Mg V -- Mg X.
The 9 -- 11 $\AA$ wavelength region contains the
 L$\beta$ line from Ne X plus the H- and 
He-like lines from Mg near 8.4 and 9.15 - 9.35 $\AA$, respectively.
The RRC of Ne IX is near 10.35 $\AA$.
The 5 -- 7 $\AA$ wavelength region contains the n=1-2 lines from both 
H-like and He-like Si, near 6.2 $\AA$ and 6.7 $\AA$, respectively.
The Mg XII L$\beta$ line is apparent at 7.15 $\AA$.  
Inner shell fluorescence lines from L shell ions of Si:  
Si VII -- Si XII are between 6.85 -- 7.05.  Lines due to 
H- and He-like S are at 4.74 and 5.05 $\AA$, respectively.  

The fluxes in table \ref{table2} can be compared with those 
given by previous studies of the NGC 1068 X-ray spectrum, by \citet{Kink02} 
and \citet{Ogle03}.  There is incomplete overlap between the lists of detected 
lines by us and these authors. They detect 40 and 60 lines, 
respectively, which can be compared with 86 in our table \ref{table2}.  Of 
these, some of the lines detected using the $XMM-Newton$ RGS by \citet{Kink02} are 
outside of the spectral range of the $Chandra$ HETG. 
There are additional discrepancies in the 
significance and identification of a small number of weak lines when 
compared with \citet{Ogle03}, though none of these is highly statistically 
significant in either our spectra or theirs.  Since the \citet{Ogle03} 
study was based on a much shorter observation than ours, it is not 
surprising that we detect a larger number of lines.  We find
general  consistency between the fluxes in table \ref{table2} and those 
of \citet{Ogle03}.  There are significant discrepancies in the 
comparison between our line fluxes and those of  \citet{Kink02}, generally 
in the sense that the  \citet{Kink02} are greater than ours by factors 
of several, possibly due to the different characteristics
of the $XMM-Newton$ telescope and the spatial extent of the NGC 1068 
X-ray emission which allows flux from a more extended region to enter the 
spectrum.


\subsection{Line wavelengths}

Figure \ref{figvoff} and Table \ref{table2} show the Doppler shifts of the line centroids for the 
strongest lines in Table \ref{table2}.  The shift is given in velocity units 
and is measured relative to the rest frame of the AGN.  Laboratory wavelengths 
are taken from the {\sc xstar} database \citep{Baut01}; most are from NIST 
\footnote{http://physics.nist.gov/PhysRefData/ASD/}.  Error 
bars are from the  errors provided by the automated Gaussian fitting procedure described 
in  the previous section.  
Wavelengths are only plotted if the errors on the wavelength are less 
than 2000 km s$^{-1}$.
Most wavelengths are consistent with a Doppler shift in the range 
400 -- 600 km s$^{-1}$.  Exceptions correspond
to RRCs, eg. Ne IX 10.36 $\AA$, for which the tabulated 
wavelengths are less precise, and weak lines such as O VIII 1s -- 5p
which is blended with Fe XIX near 14.8 $\AA$.
This figure shows that there is  evidence for wavelength shifts which are 
larger than the errors on the line centroid determination.   Examples of this 
include the O VIII L$\alpha$ line compared with the 1s$^2$ -- 1s2s$^3S$ ($f$)
 line of O VII.

Figure \ref{figvoff}, lower panel,  and Table \ref{table2}
show the Gaussian widths of the lines for the 
strongest lines.  The width is given in velocity units, 
i.e. $\sigma_{\varepsilon} c/\varepsilon$ 
where $\sigma_{\varepsilon}$ is the Gaussian full width at half maximum 
in energy units and $\varepsilon$ 
is the line energy.
Error bars are from the confidence errors provided by our automated 
fitting procedure as described above.  
Wavelengths are only plotted in figure \ref{figvoff} 
if the errors on the wavelength are less 
than 2000 km s$^{-1}$.
This shows that there is no single value for the width which 
is consistent with all the lines.    Examples of broader lines 
include the O VIII L$\alpha$ line, and contrast with 
the forbidden  line $f$ of O VII, for which the width is smaller.
The full width at half maximum which fits to most lines is approximately 1400 km s$^{-1}$.
This corresponds to a Gaussian sigma of $\simeq$ 600 km s$^{-1}$ which 
we adopt in our numerical models discussed below.

\subsection{Spatial Dependence}
\label{spatialsec}

The discussion so far has utilized data extracted using the standard 
size extraction region from the HETG, which has width 1.3 $\times 10^{-3}$ degrees or 
4.8 arcsec.  We have also explored possible spatial dependence of the spectrum
in the direction perpendicular to the dispersion direction by extracting 
the data using regions which are half and double the angular size, i.e. 2.4 and 9.6 arcsec 
wide. We then carry the fitting steps described above.     
Figure \ref{fig7} illustrates
the positions of these regions superimposed on the zero order image.

In this way we can study the influence of adding successive regions further 
from the dispersion axis to the spectrum.   Figure \ref{fig8} shows the 
ratios of the line fluxes in the standard extraction region (green hexagons) 
and the extraction double the standard size (red diamonds) compared 
to the line flux in the extraction region half the standard width. 
 This shows that the line emission outside half width region 
is significant; the values of the ratios of essentially all the lines 
are greater than unity, and many are $\sim$2.  
Plus, the emission from outside 
the standard 4.8 arcsec extraction region is not negligible;
the values of the ratio for this region (red symbols) are on average 
greater than the values of the ratio for the standard region.    
The principal difference between standard region and the double size region
is the importance of the radiative excitation emission, which is larger 
in the standard region.  This is consistent with results from 
\citet{Ogle03}.  For example, for the He-like lines of Ne IX, the G ratio is
larger in the standard region spectrum than in the spectrum from the double size region.  For the O VII lines,
the G ratios are similar for the two regions, but the R ratio is lower in 
the standard region, implying higher density there.

These results demonstrate that most of the line emission is contained 
within the standard extraction region, 4.8 arcsec wide, and 
most line ratios are unaffected when emission from outside the standard 
extraction region is included.  
It is worth noting that the physical length scale corresponding to 
this angular size is approximately 1 arcsec = 72 pc \citep{Blan97}, 
so the standard extraction region corresponds 
to a total size of 347 pc, or a maximum distance from the 
central source of approximately 174 pc.  This is much larger than the region 
where the broad line region and the obscuring torus are likely to lie, 
which is $\simeq$ 2 -- 3 pc \citep{Jaff04}.
For the remainder of this paper we will 
discuss primarily analysis of data from the standard extraction region. 

%
%
\subsection{He-like lines}
\label{helines}

The He-like lines in Table \ref{table2} provide sensitive diagnostics
of emission conditions:  excitation mechanism, density, and ionization balance.
These challenges of fitting these lines are apparent from  the region containing 
the H and He-like lines from the elements O, Ne, Mg, Si, and S.  This shows the
characteristic three lines from the n=2 -- n=1 decay of these ions:  the resonance ($r$), 
intercombination ($i$) and forbidden ($f$).  These lines provide useful diagnostics 
of excitation and density.  These are described in terms of the ratios $R=f/i$ and $G=(f+i)/r$
\citep{Gabr69a,Gabr69b}.
$R$ is indicative of density, since the energy splitting between the upper levels of the 
two lines (2$^3$S and  2$^3$P) is small compared with the typical gas temperature and 
collisions can transfer ions from the 2$^3$S to the  2$^3$P when the density is above a 
critical value.  $R$ can also be affected by radiative excitation from the 1s2s$^3$S to the 1s2s$^3$P 
levels, but this requires photon field 
intensities which are greater than what is anticipated for NGC 1068.  
In our modeling we include this process, using an extension of the global 
non-thermal power law continuum, and find that it is negligible.  We do not 
consider the possibility of enhanced continuum at the energy of the 
2$^3$S to the  2$^3$P provided by, eg., hot stars.
$G$ is an indicator of temperature or 
other mechanism responsible for populating the various $n=2$ levels.  At high 
temperature, or when the upper level of the $r$ line (2$^1$P) is excited in some other way, 
then $G$ can have a value $\leq$1; in the absence of such conditions, the  2$^3$S and  2$^3$P
levels are populated preferentially by recombination and $G$ has a characteristic 
large value, $G\geq4$.   Table \ref{table2}  shows that the various elements have common values of $R$, 
in the range 1 -- 3, indicating moderate density.  More interesting is the fact that 
there is a diversity of values for $G$:  O and Ne have values 2.5 -- 3 while the other 
elements all have smaller values, $G\leq$ 1.5.  

The $G$ ratio is most affected by the process responsible for excitation of 
the $r$ line.  This can be either electron impact collisions or radiative 
excitation.  Collisions are less likely in the case of NGC 1068 due to the 
presence of radiative recombination continua (RRCs) from H-like and He-like 
species in the observed spectra \citep{Kink02}. 
This indicates the presence of fully stripped and H-like ions, 
which produce the RRCs, but at a temperature 
which is low enough that the RRCs appear narrow.  Typical RRC widths in NGC 1068 
correspond to temperatures which are $\leq 10^5$K. A coronal plasma, in which 
electron impact collisions are dominant, would require a temperature $\geq 10^6$ K
in order to produce the same ions.  Radiative excitation can produce $G$ ratios as small 
as 0.1.  Thus, the $G$ ratios for NGC 1068 indicate the importance of radiative excitation 
for Mg and Si, while O and Ne are dominated by 
recombination.   

Radiative excitation depends on the presence of strong unattenuated continuum 
from the central source in order to excite the $r$ line.  A consequence of the 
large cross section for radiative excitation in the line is the fact that the line 
will saturate faster than continuum.  At large column densities from the source, 
the radiation field is depleted in photons capable of exciting the $r$ line, while 
still having photons capable of ionizing or exciting other, weaker lines.  Thus, 
the $G$ ratio is an indicator of column density.  He-like lines formed 
after the continuum has traversed a large column density 
will have large $G$ ratios, corresponding 
to primarily recombination.  He-like lines formed after the continuum 
has traversed a small column density will have small $G$ ratios, corresponding 
to radiative excitation \citep{Kink02}.  

Figure \ref{fig3a} and \ref{fig3b} display the G ratio for the elements O, Ne, Mg, and Si from the 
NGC 1068 HETG observations.  These are shown with error bars as the 
red crosses in the centers of the various frames.  
This shows that the ratios differ between the elements:  
O and Ne have G$\simeq$2 -- 3, while Si  and Mg have G$\simeq$1.  
This differing behavior suggests that the He-like O and Ne lines are emitted by 
recombination, while the He-like lines of Mg and Si are emitted by radiative excitation.   
Since radiative excitation is suppressed by large resonance line 
optical depths, this suggests small optical depths in Si and Mg and larger optical depths in the 
lines from other elements.  The differences in optical depths are significant; as shown by 
\citet{Kink02} equivalent hydrogen columns of 10$^{23}$ cm$^{-2}$ or more are required to suppress 
radiative excitation.

Also plotted in figures \ref{fig3a} and \ref{fig3b} are the values of these ratios produced by photoionization models 
consisting of a single slab of gas of given ionization parameter at the illuminated face and 
given column density.  The solid curves correspond to constant total slab column density,  
where black= 10$^{23.5}$, blue=10$^{22.5}$, green=10$^{21.5}$ and red=10$^{20.5}$.
The dashed curves correspond to constant 
ionization parameter in the range 1$\leq$log($\xi$)$\leq$3.
Here the ionization parameter is $\xi = 4\pi F/n$ where $F$ is the ionizing energy 
flux in the 1 -- 1000 Ry range and $n$ is the gas number density.
The role of radiative excitation is apparent from the fact that the low
column density models (red curves) produce smaller $G$ values than the high 
column density models (black curves).

The results in  figures \ref{fig3a} and \ref{fig3b} show that no single 
value of the photoionization model parameters can 
simultaneously account for all the ratios. O and Ne require log($\xi$)$\simeq$1 and 
column$\geq 10^{23}$cm$^{-2}$, while Si requires log($\xi$)$\simeq$2 and 
column$\leq 10^{22}$cm$^{-2}$ and Mg is intermediate in both $\xi$ and column.


Also plotted in figures \ref{fig3a} and \ref{fig3b} are values for certain other ratios, in the same 
units as for the He-like lines, and also including the model values.  
These include the higher series allowed lines from some He-like 
ions, eg. 1-4/1-2 vs 1-3/1-2 for O VII, where 1-2 includes the $r$, $i$, and $f$ 
lines while 1-4 and 1-3 include only the resonance component.  These ratios depend
primarily on ionization parameter and only very weakly on column density. 
This is because recombination cascades make larger values of both ratios 
than radiative excitation, so recombination tends to dominate production of these 
lines.  

\subsection{Iron K Line Region}
\label{feksection}

The iron K lines from NGC 1068 have been the subject of previous studies by \citet{Matt04, Iwas97},
who showed that the line consists of three components, corresponding to near neutral, and 
H- and He-like ion species.  The HETG provides higher spectral resolution than these 
previous studies, but fewer total counts.  
Figure \ref{fig4} shows the spectrum in the region between 1.5 -- 2.5 $\AA$ 
and reveals the three components of the iron K lines.  The most 
prominent are a feature at 6.37 keV (1.95 $\AA$), consistent with neutral or near-neutral iron, 
a feature at 6.67 (1.85 $\AA$) indicative of He-like iron, and a feature near 7 keV (1.75 $\AA$) which 
may be associated with a combination of the K absorption edge from near-neutral gas and emission from 
H-like iron.  \cite{Matt04} also separately identify emission from Be-like iron; 
we are not able to make such an identification.  We do find emission at other nearby wavelengths, 
but this is suggestive of a broad component or blended emission from various other species. 

We model the Fe K line region using `analytic' models which utilize the {\sc xstar} database 
and subroutines.  These include optically thin models which include radiative excitation 
(which we denote `scatemis') and models which do not include this process
(`photemis'), plus power law continuum.  Both these models are available for use as 
`analytic' models in {\sc xspec} from the {\sc xstar} site
\footnote{http://heasarc.gsfc.nasa.gov/docs/software/xstar/xstar.html}.
Radiative excitation cannot excite the neutral line, and we model it as 
emission from gas at log($\xi$)=-3.  The emission component with radiative excitation
has log($\xi$)=3, corresponding to highly ionized gas in which 
the H- and He-like stages are the most abundant ionization stages
of Fe.  In the HETG data the He-like triplet is not fully resolved.
Nonetheless, the centroid of the He-like component is closer to the 
wavelength of the Fe XXV $r$ line than it is to the wavelength of the $f$
or $i$ lines, so we find a better fit with a resonance excitation model.
The fit is shown in figure \ref{fig4}.  The best-fit 
parameters are summarized in table \ref{table5}; we find $\chi^2$=103 for 206
degrees of freedom for the 1 -- 2.5 $\AA$ (5 - 12.4 keV) spectral region.
The flux in the neutral-like line is  3.8 $^+_-0.44 \times 10^{-5}$ cm$^{-2}$ s$^{-1}$. 

 We also point out 
what appears to be unresolved emission in the wavelength region 
near $\simeq$ 1.9 $\AA$, between  the He-like line and the neutral-like 
line.  We have not attempted to model this.   It may be due to a broadened
component of the modeled lines, or an unresolved blend from other high 
ionization stages of iron.  These could include L shell ions between Li-like
and Ne-like.  
The two ionization parameters which we use to fit the iron K lines are both 
higher and lower than the values used to fit the remainder of the emission 
spectrum, described in the next section.  The high ionization parameter, 
log($\xi$)=3, may be an extension of the ionization parameter range needed to fit 
the lines in the energy band below $\sim$ 5keV.  The low ionization parameter, 
log($\xi$)=-3, likely corresponds to the torus.  Neither of these components 
produces significant emission below $\sim$ 5keV, so we will not discuss  them further 
in the model fits described in the following section.

The iron K line complex contains more flux than any other line in the spectrum.
This is consistent with the large normalization for the low ionization component 
that is required to fit the spectrum.  This was discussed by \citet{Krol87,Nand06}.
We have not attempted to fit for a Compton shoulder on the 
on the red wing of the neutral-like 
iron line, although such a component may be present in figure \ref{fig4} \citep{Matt04}.  
We also test for the presence of Ni K$\alpha$ in our spectrum
and find an upper limit of 1.5 $\times 10^{-5}$ cm$^{-2}$ s$^{-1}$.  This is 
less than the flux for Ni K$\alpha$ claimed by \citet{Matt04}, which was 
5.6$^{+1.8}_{-1.0} \times 10^{-5}$  cm$^{-2}$ s$^{-1}$.  One possible 
explanation could be that their data was extracted from a much larger
spatial region, $\simeq$40 arcsec, and is also affected by 
a nearby point source.

\subsection{Iron L lines}
\label{fellines}

The $Chandra$ HETG observation allows study of the lines from the L shell 
ions of iron, Fe XVII -- XXIV, in more detail than has previously 
been possible.  Table \ref{table2} and figures \ref{fig1a} --  \ref{fig1d} show these 
in the wavelength range 10 -- 17 $\AA$.  
Notable features include the well known `3C' and `3D' \citep{Park73} lines of Fe XVII at 
15.01 and 15.26 $\AA$
(826 and 812 eV), respectively.  The intensity ratio of these lines is a 
topic of interest in the study of coronal plasmas; we find a ratio of 
2.8 $^{+0.8}_{-0.5}$.  This is consistent with several other lab measurements 
and astrophysical observations.  It also reflects the current apparent 
discrepancy with calculations, which generally produce values of this ratio 
which are 3.5 or greater \citet{Bern12}.    The {\sc xstar} models, described 
below, reflect this situation and produce a value for this ratio 
which is $\simeq$3.

Other strong lines in the spectrum include the Fe XVII -- XVIII complex between
11 -- 11.5 $\AA$, Fe XXII 11.77 $\AA$.  Fe XVII 12.12 $\AA$ is blended with 
Ne X L$\alpha$. The Fe XVII 2p -- 4d line at 12.26  $\AA$ is apparent, 
but the {\sc xstar} models with radiative excitation (described in more detail 
in the following section) fail to produce as much flux as is observed. 
Fe XX 12.58  $\AA$ is well fit, as is Fe XX 12.81, 12.83  $\AA$.  The 13.6  $\AA$
blend contains both Ne IX and Fe XIX 2p -- 3d lines, produced primarily 
by radiative excitation.   Also contributing at 13.82 $\AA$ is the 
Fe XVII 2s -- 3p line; this line can only be produced efficiently by radiative excitation 
and it is clearly apparent on the red wing of the Ne IX $f$ line.
The 2p-3d lines of Fe XVIII near 14.2 and out to 14.6  $\AA$ are fitted by the models.
Longward of 15  $\AA$ the most notable feature is the Fe XVII line 
at 17.1  $\AA$, which is an indicator of recombination. 
 
Also plotted in figure \ref{fig3b} is the ratio 2p-4d/2p-3s vs 2p-3d/2p-3s for the Ne-like Fe XVII.
The latter ratio is sensitive to the effects of radiative excitation since 2p-3d has 
a higher oscillator strength than 2p-3s, while 2p-3s is emitted efficiently 
via recombination cascade \citep{Lied90}.  Here it is clear from figures 
\ref{fig1a} -- \ref{fig1d} that the 2p-3d line is strong, therefore requiring 
that radiative excitation be efficient and the cloud column density 
must be low for the gas producing the Fe XVII lines.

\subsection{Fluorescence Lines}

Fluorescence lines are emitted during the cascade following K shell 
ionization.  This is a likely signature of photoionization because, 
for ions with more than 3 electrons, the rate for K shell 
ionization by electron impact by a Maxwellian velocity distribution 
never exceeds the valence shell 
cross section.  Fluorescence lines probe the existence of ions with 
valence shell energies far below the observable X-ray range, including 
neutral and near-neutral ions.   In the NGC 1068 spectrum, fluorescence 
lines include the K$\alpha$ line of iron shown in figure \ref{fig4}, plus 
a series of Si lines near 6.8 $\AA$.  These correspond to L shell ions 
Si, likely Si VIII -- X.  The conditions under which these ions are likely 
to be abundant, i.e. the ionization parameter in a photoionization equilibrium
model, are not very different from those corresponding to N VII and O VII, the 
lowest ionization species in the spectrum.  Thus they do not 
provide additional significant insight into the existence or quantity 
of material at low ionization in NGC 1068.  

\section{Model}
\label{model}
\label{modelsec}

A goal of interpreting  the X-ray spectrum of NGC 1068 is to understand 
the distribution of gas: its location relative to the central continuum 
source, its velocity, density, element abundances, temperature, ionization 
state.  Fitting of models which predict these quantities to the data are 
the most effective way to derive this information, although 
there is a range of levels of detail and physical realism for doing this.
We have examined this through the use 
of photoionization table models in {\sc xspec}, which are described in the Appendix.

Based on what was presented in section \ref{data} we can summarize some of the 
criteria for a model for the NGC 1068 HETG spectrum: (i) The model must 
account for the H- and He-like line strengths from abundant elements from N to Ca
(we exclude the iron region from this discussion); (ii) The model must account for 
the G ratios from the elements O, Ne, Mg, Si, which have G$\simeq$3 for O and Ne but
G$\simeq$1 for Mg and Si; (iii) The model must account for 2p-3d/2p-3s ratio in 
Fe XVII which is indicative of recombination rather than radiative excitation; 
(iv) The model must account for the strengths of the RRCs, which have strengths
comparable to the corresponding resonance lines for O VII.  From criterion (i) 
it is clear that the emitting gas must have a range of ionization parameter; 
no single ionization parameter can provide these ions.  From criterion (ii) it is 
likely that a range of gas column densities are needed.  The hypothesis that 
the lower ion column densities associated with lower abundance elements (Mg and Si) 
can account for the diversity in $G$ ratios can be tested in this way.  Criterion 
(iv) implies that some of the gas must have high column in order to have a detectable
contribution from recombination.

Our fit uses three components with varying ionization parameters; 
these are spread evenly between log($\xi$)=1 and log($\xi$)=2.6.
The HETG spectrum is relatively insensitive to gas outside this range of ionization parameters, 
except for the iron K lines.  Given the strength of both the neutral-like and the H- and He-like 
iron K lines, it is plausible that there is an approximately  continuous 
distribution of ionization parameters which extends beyond the range
considered here.    The three ionization parameter components 
in our fit correspond crudely to the regions dominating the emission for the elements 
of varying nuclear charge:  the low ionization parameter component dominates the emission 
for N and O, the intermediate component dominates the Fe L lines and Ne and Mg, and 
the high ionization parameter component dominates for Mg and Si.  
For each ionization parameter, we include models with two column densities:  
3 $\times 10^{22}$ cm$^{-2}$ and 3 $\times 10^{23}$ cm$^{-2}$.  In this way the models span 
the likely ionization parameter and column density for which the strongest lines 
we observe are emitted.  We use a constant turbulent line 
width (sigma) of 600 km s$^{-1}$  which adequately accounts for the widths 
the majority of features.  Results of our fit are summarized in table \ref{table4}.

\begin{figure}[ht]
\includegraphics[angle=0, scale=0.8]{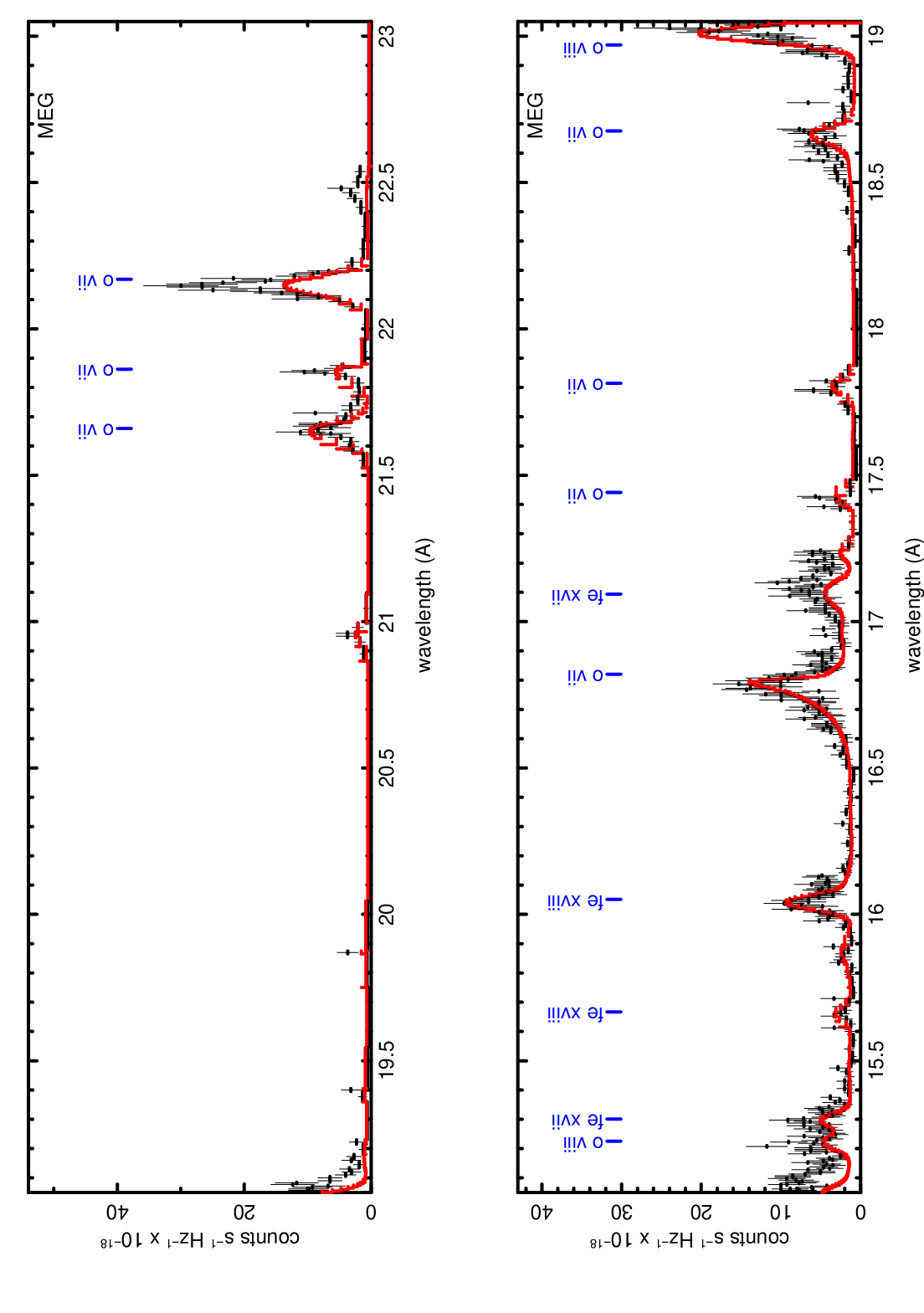}
\caption{\label{fig1a} Spectrum showing fits to table model described 
in the text.  Vertical axis is counts s$^{-1}$Hz$^{-1}$ scaled according to the maximum flux 
in the panel.}
\end{figure}

\begin{figure}[ht]
\includegraphics[angle=0, scale=0.8]{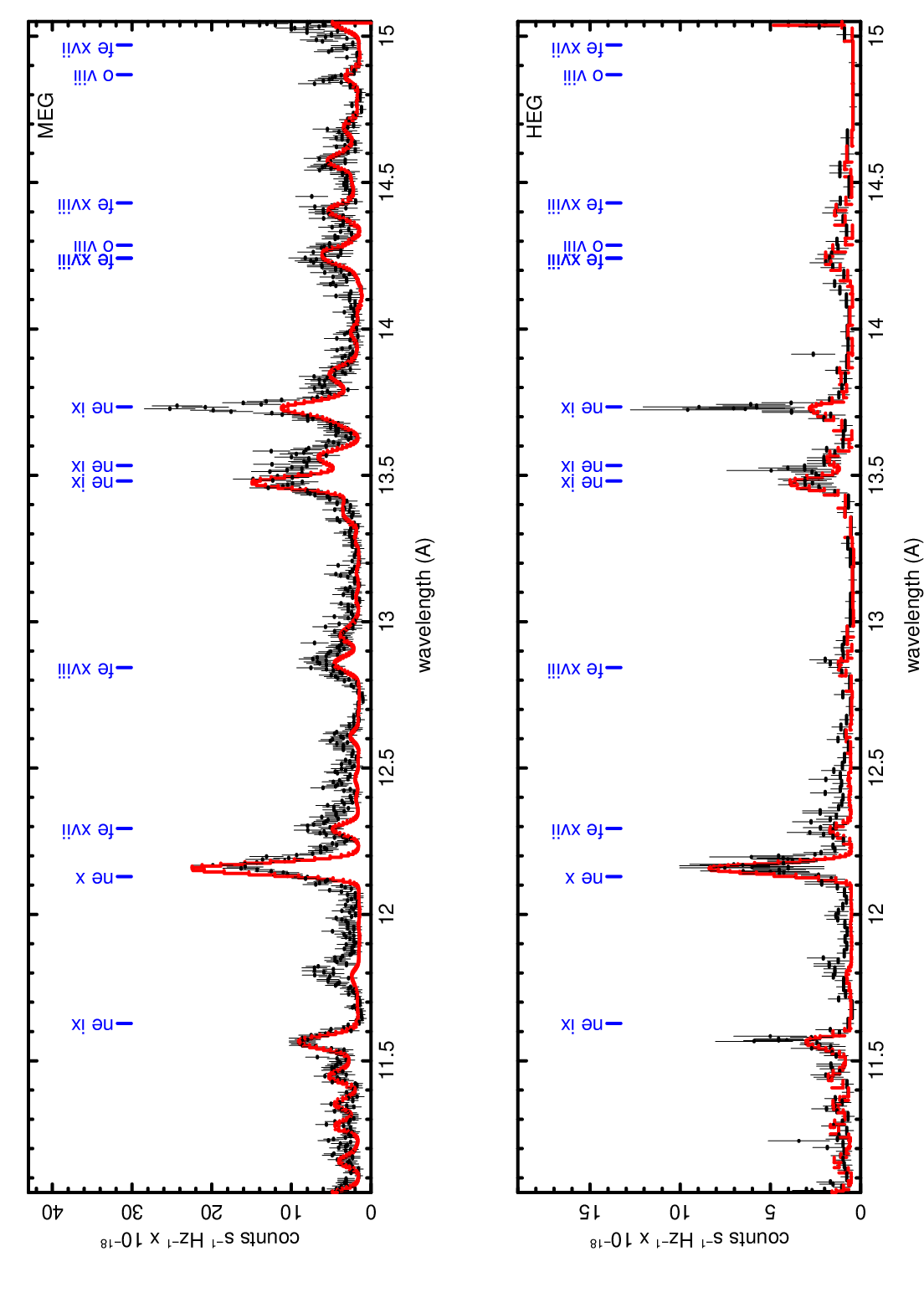}
\caption{\label{fig1b} Spectrum showing fits to table model described 
in the text.  Vertical axis is counts s$^{-1}$Hz$^{-1}$ scaled according to the maximum flux 
in the panel.}
\end{figure}

\begin{figure}[ht]
\includegraphics[angle=0, scale=0.8]{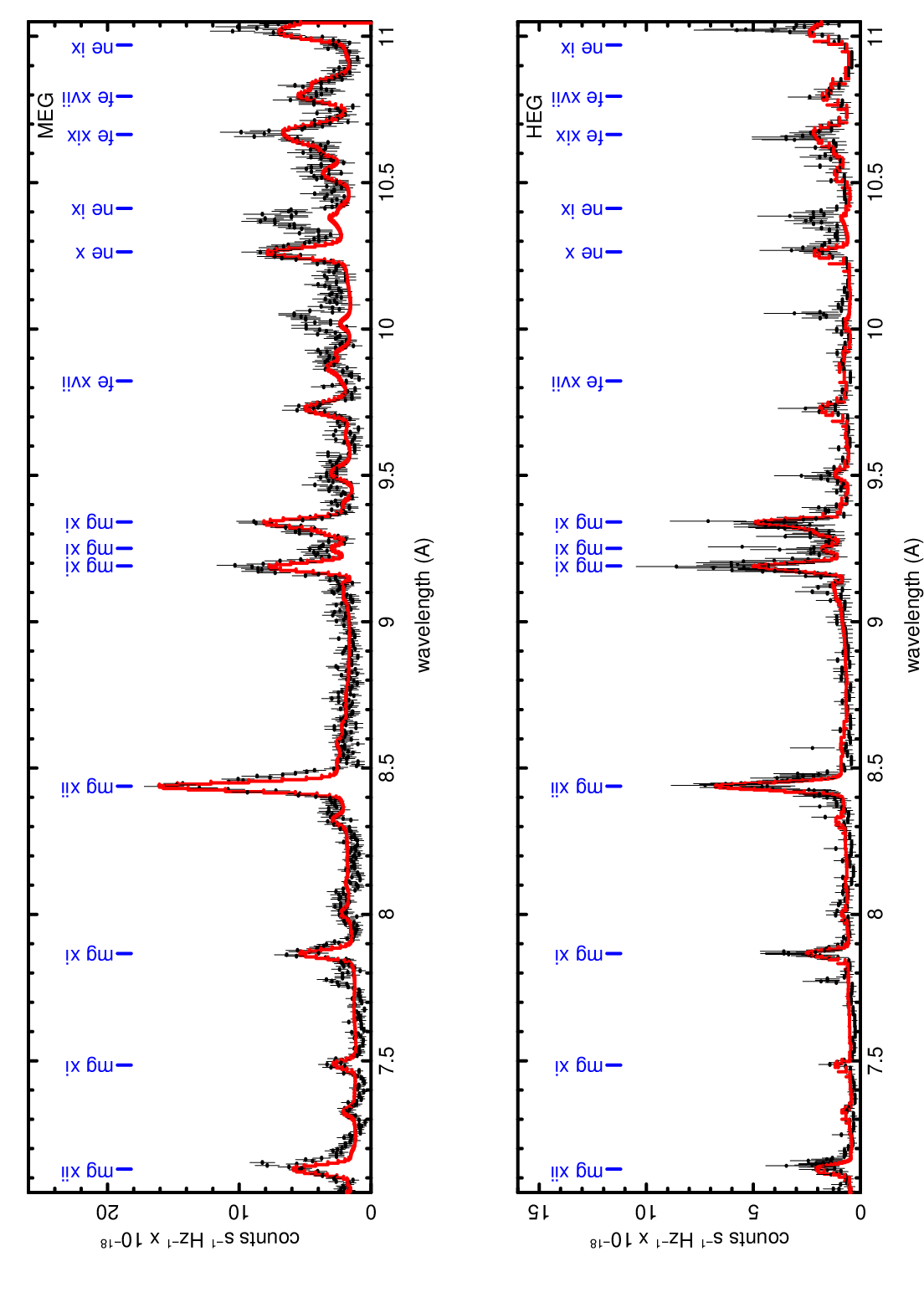}
\caption{\label{fig1c} Spectrum showing fits to table model described 
in the text.  Vertical axis is counts s$^{-1}$Hz$^{-1}$ scaled according to the maximum flux 
in the panel.}
\end{figure}

\begin{figure}[ht]
\includegraphics[angle=0, scale=0.8]{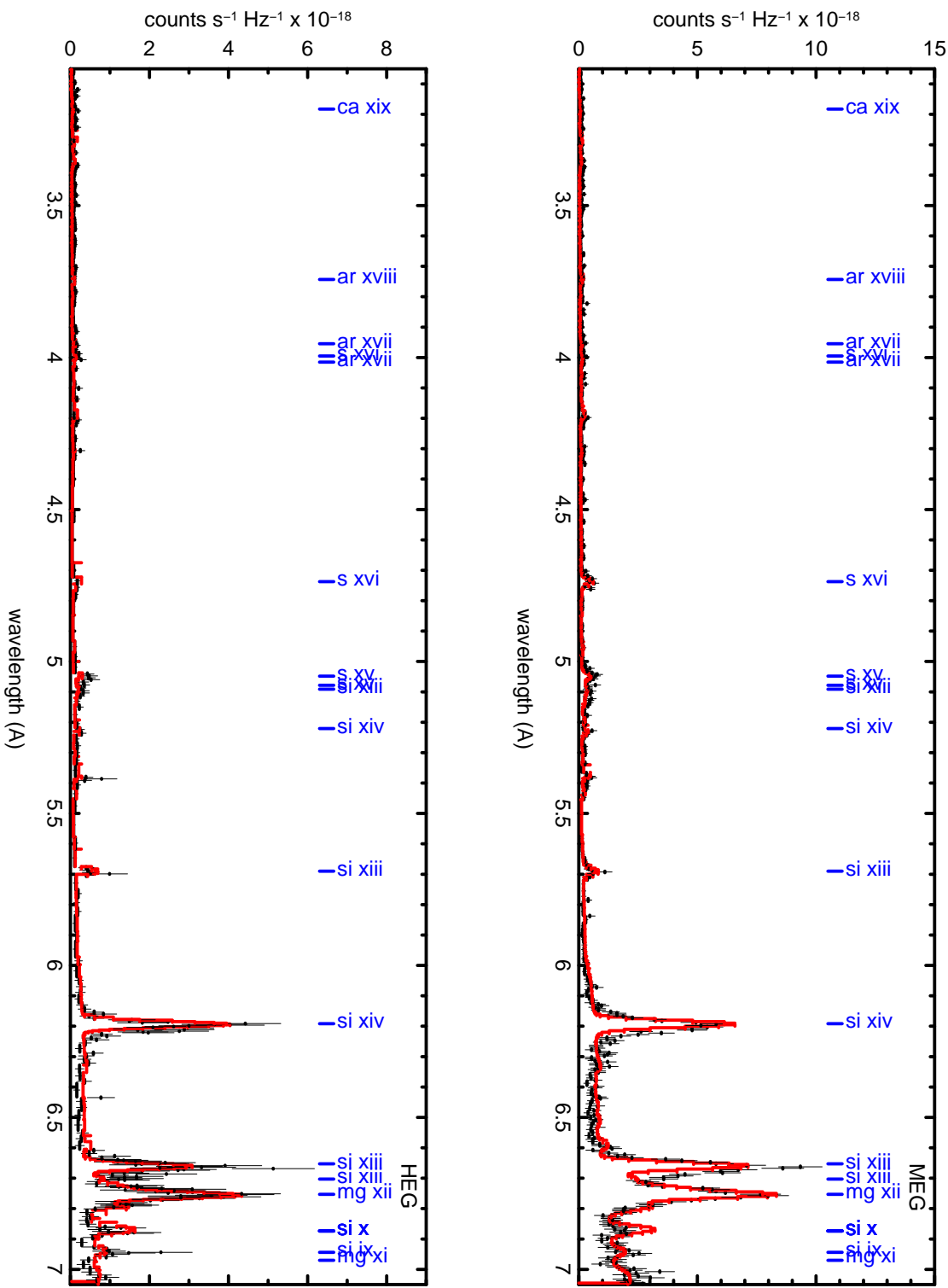}
\caption{\label{fig1d} Spectrum showing fits to table model described 
in the text.  Vertical axis is counts s$^{-1}$Hz$^{-1}$ scaled according to the maximum flux 
in the panel.}
\end{figure}

\begin{figure}[ht]
\includegraphics[angle=0, scale=0.8]{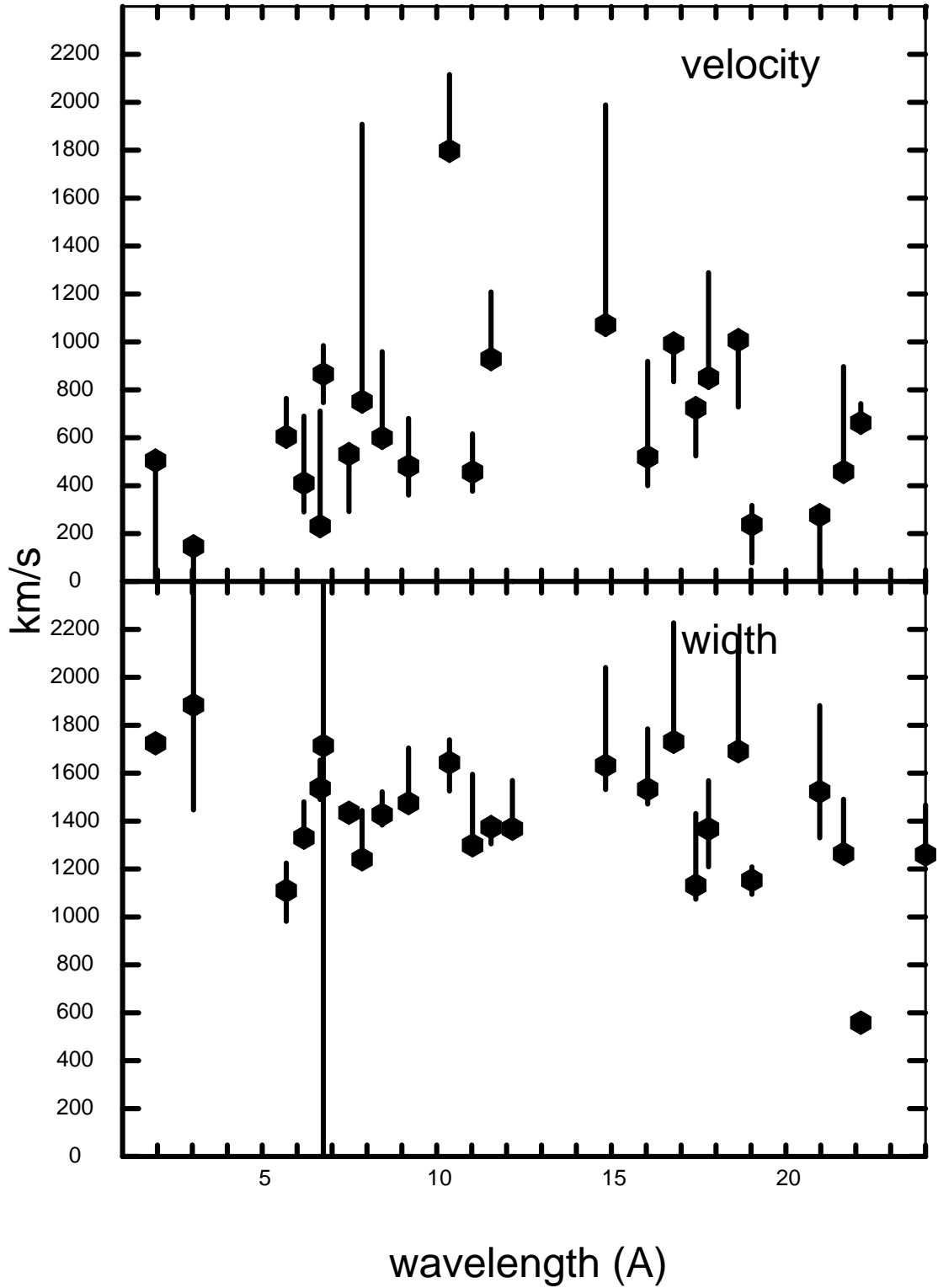}
\caption{\label{figvoff}  Plot of the widths and velocity 
offsets of the lines shown in table \ref{table2} in 
km s$^{-1}$.  Only lines for which errors on the line 
width can be derived are plotted.}
\end{figure}

\begin{figure}[ht]
\includegraphics[angle=0, scale=0.7]{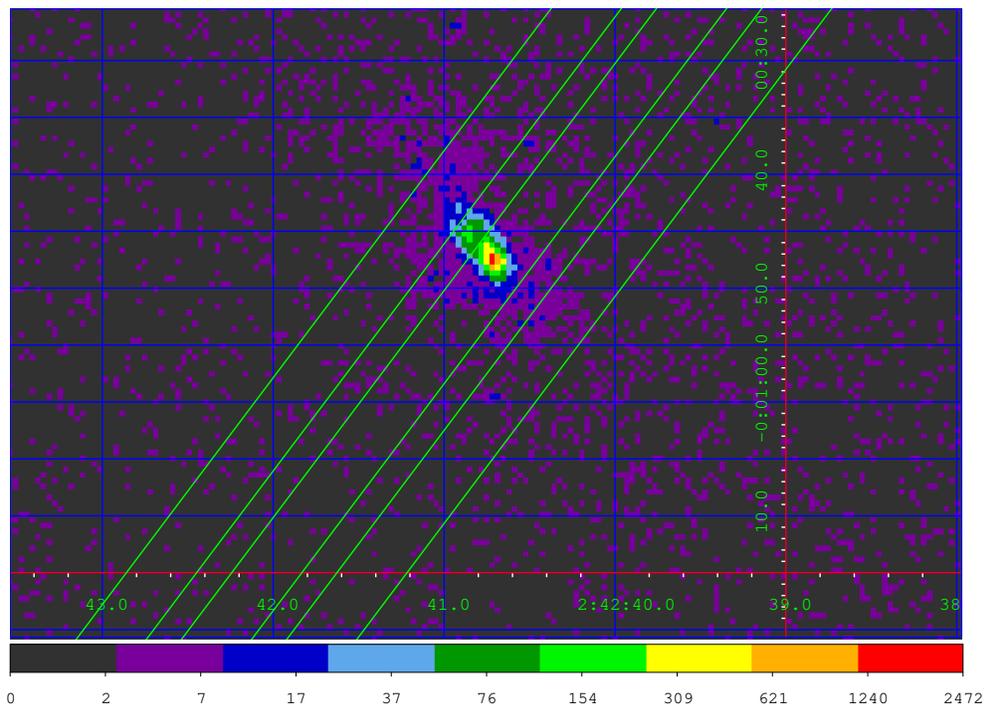}
\caption{\label{fig7} Zero order image with positions of extraction regions shown.
Lines corresponds to standard extraction region (4.8 arcsec width) plus half 
and double size regions as discussed in text.  Only heg arm is shown for clarity.}
\end{figure}

\begin{figure}[ht]
\includegraphics[angle=0, scale=0.6]{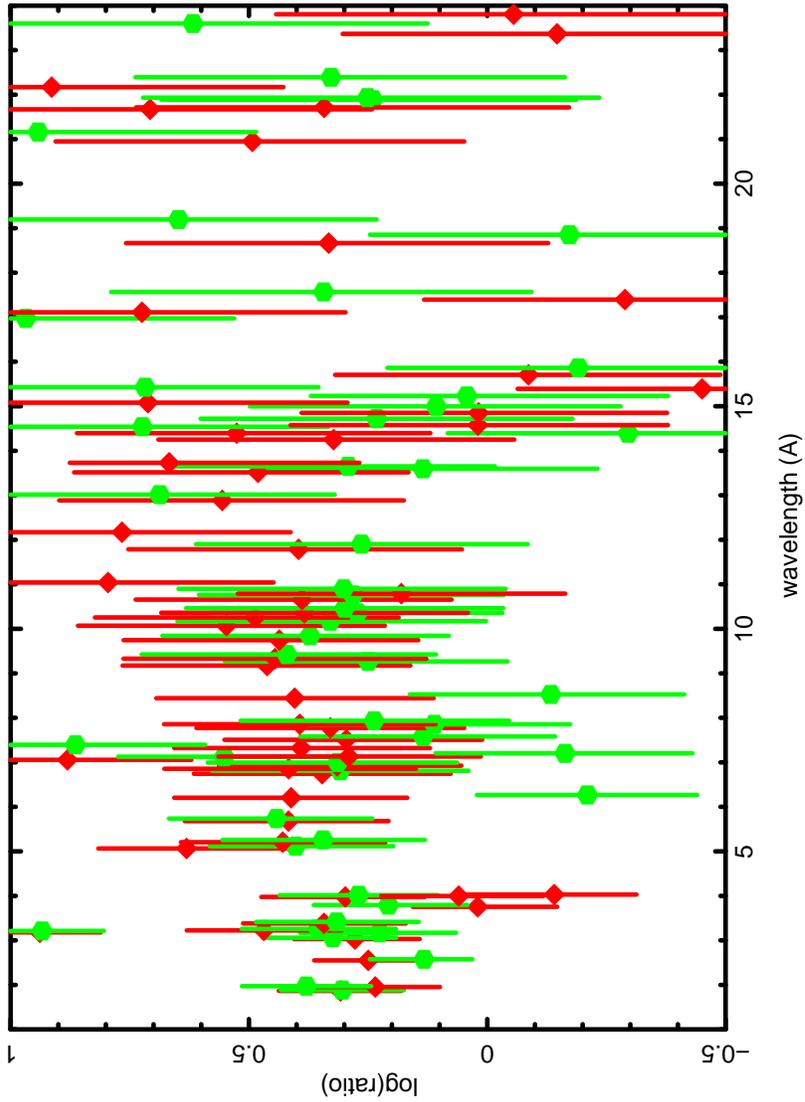}
\caption{\label{fig8} Values for the log of the ratios of line fluxes
for regions with the standard size (4.8 arcsec) (green 
hexagons) and double this value (red diamonds) 
compared with line fluxes for a region half the standard size.}
\end{figure}

\begin{figure}[ht]
\includegraphics[angle=0, scale=0.7]{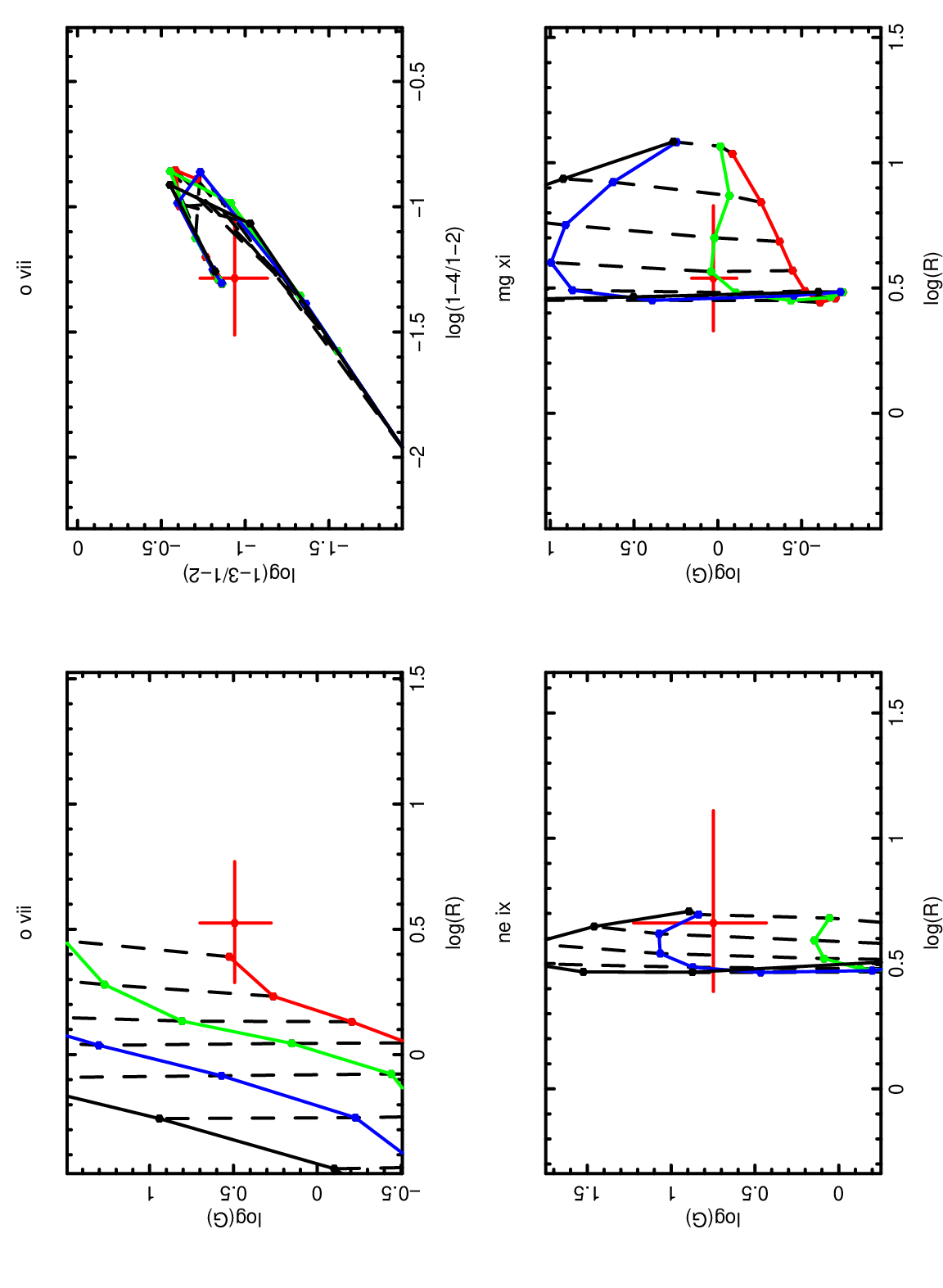}
\caption{\label{fig3a}  Plot of the locus of points in the plane 
of the He/H line ratio vs. the G-ratio.  See the text for definitions.
The solid curves correspond to constant column density in 
where black= 10$^{23.5}$, blue=10$^{22.5}$, green=10$^{21.5}$ and red=10$^{20.5}$.
The dashed curves correspond to constant 
ionization parameter in the range 1$\leq$log($\xi$)$\leq$3.
 Red bars denote the range of measured values.}
\end{figure}

\begin{figure}[ht]
\includegraphics[angle=0, scale=0.7]{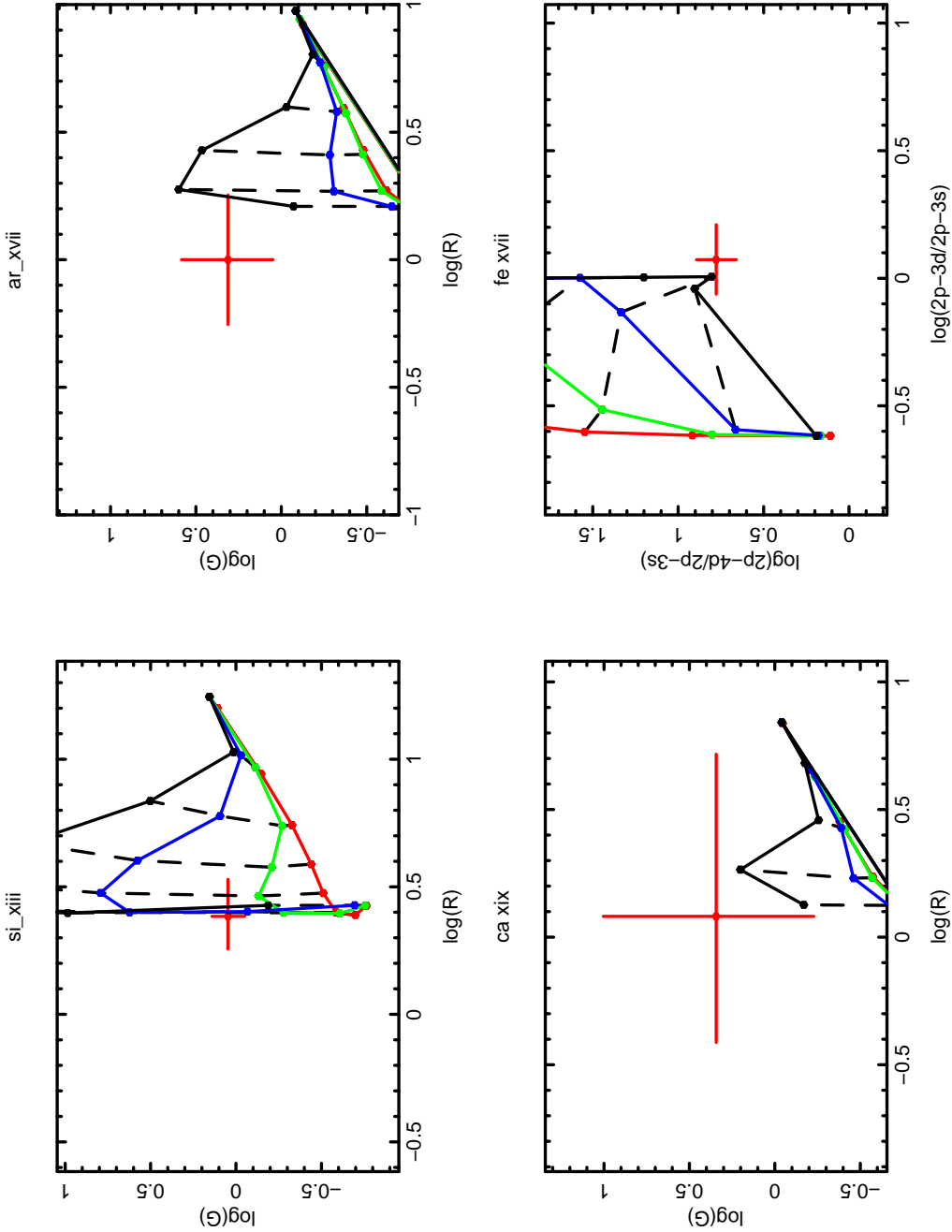}
\caption{\label{fig3b}  Plot of the locus of points in the plane 
of the He/H line ratio vs. the G-ratio.  See the text for definitions.
The solid curves correspond to constant column density in 
where black= 10$^{23.5}$, blue=10$^{22.5}$, green=10$^{21.5}$ and red=10$^{20.5}$.
The dashed curves correspond to constant 
ionization parameter in the range 1$\leq$log($\xi$)$\leq$3.
 Red bars denote the range of measured values.}
\end{figure}

\begin{figure}[ht]
\includegraphics[angle=0, scale=0.6]{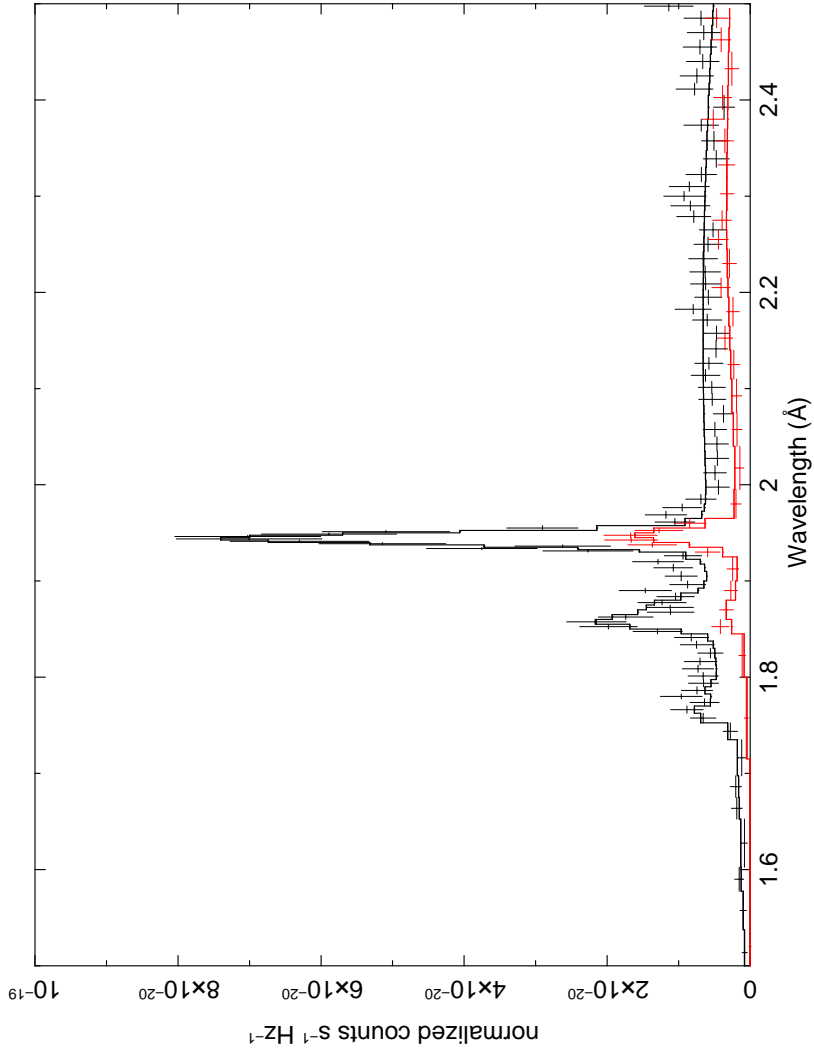}
\caption{\label{fig4}  Fe K line region. Lines are apparent from 
near-neutral Fe near 1.95 $\AA$, He-line near 1.85 $\AA$ 
and H-like near 1.75 $\AA$.  The model consists of components with 
and without radiative excitation from the analytic model described 
in section \ref{modelsec}, plus power law continuum. }
\end{figure}

The red curves in figures  \ref{fig1a} -- \ref{fig1d} 
show that this model fits the strengths of almost all the strong features 
in the spectrum.  
This fit adopts a net redshift for the line emitting gas of 0.0023, corresponding 
to an outflow velocity of 450 km s$^{-1}$.  This is marginally less than  typical 
speeds from UV and optical lines of $\leq$600 km s$^{-1}$ (both blue- and red-shifted) \citep{Cren00},
but is consistent with the velocities plotted in figure \ref{figvoff}.

\begin{figure}[ht]
\includegraphics[angle=0, scale=0.6]{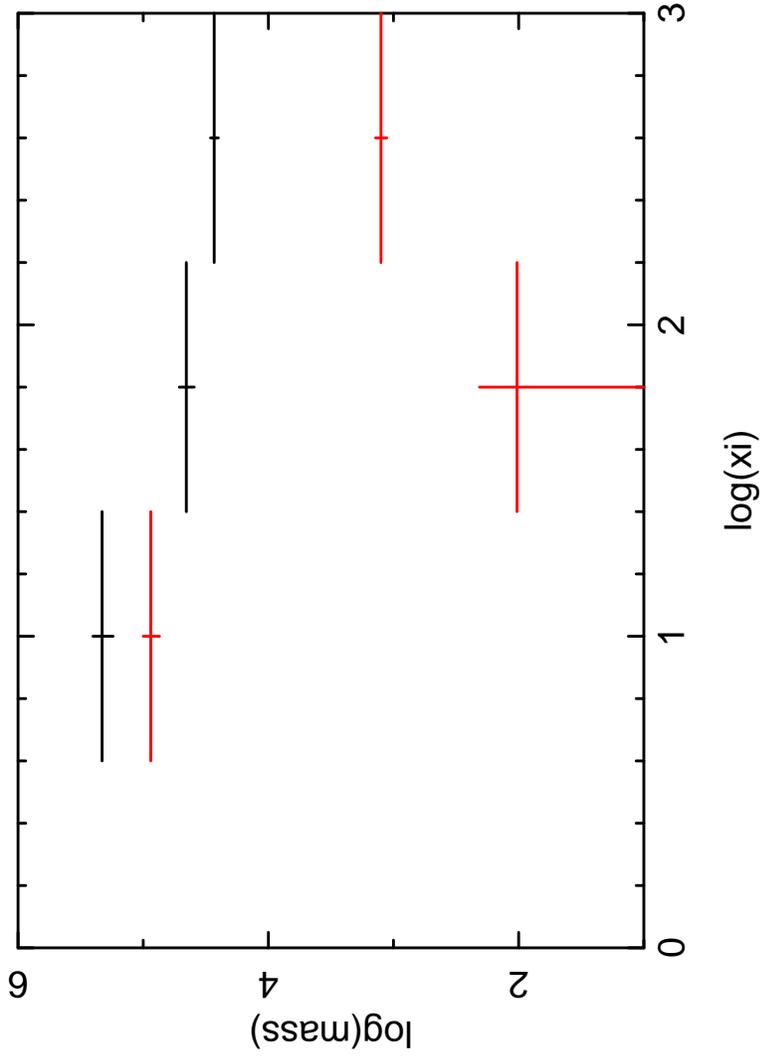}
\caption{\label{fig5}Distribution of mass among the emission components
used to model the spectrum.  Black points correspond to the 
component with column 3 $\times 10^{23}$ cm$^{-2}$ and  
red corresponds to the component with column 3 $\times 10^{22}$ cm$^{-2}$}
\end{figure}

\begin{figure}[ht]
\includegraphics[angle=0, scale=0.6]{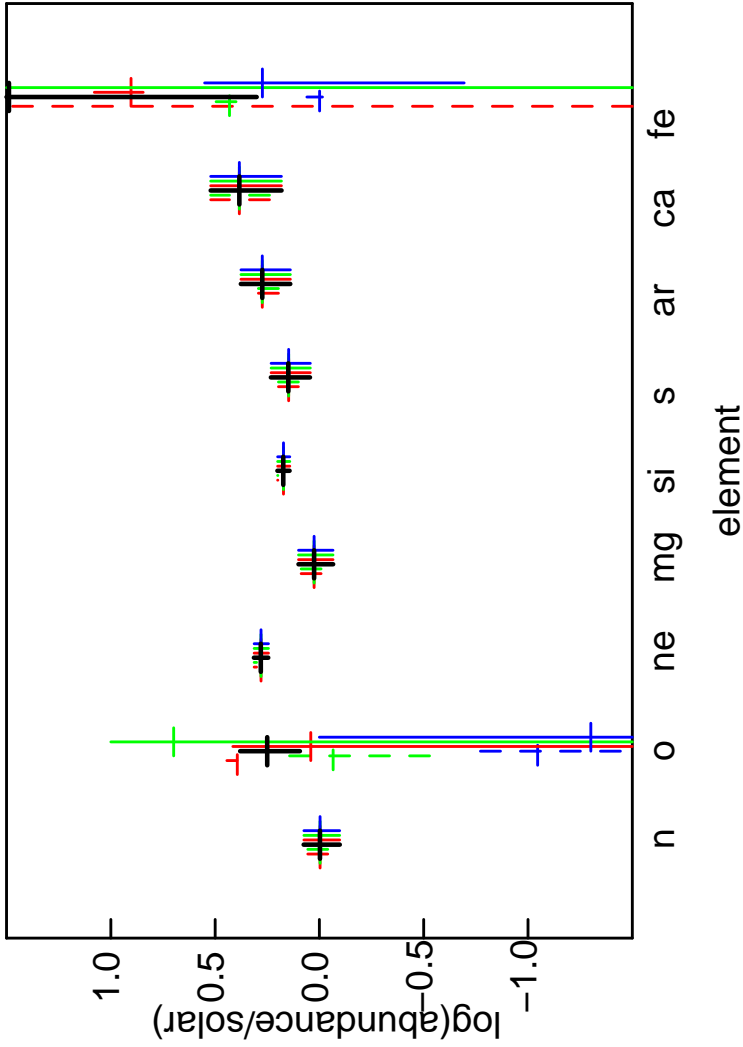}
\caption{\label{fig6} Element abundances from best-fit models.
Colors red, green and blue correspond to log($\xi$)=1, 1.8, 2.6, 
respectively; solid corresponds column 3 $\times 10^{23}$ cm$^{-2}$ 
and dashed corresponds column 3 $\times 10^{22}$ cm$^{-2}$.  Black 
points are average over the best-fit model. } 
\end{figure}

Free parameters of the fits include the elemental abundances 
and the normalizations for the six model components.  Experimentation 
shows that the fit is not substantially improved when all the abundances 
are allowed to vary independently, so we force the abundances of N, Ne, Mg, Si, S, Ar and Ca 
to be the same for all the models.  We allow the abundances of O and Fe to 
vary independently for all the models.  Our best fit is shown in figures  \ref{fig1a} -- \ref{fig1d} 
has $\chi^2/\nu=2.34$ for   2561  degrees of freedom. 
More results are shown in figures \ref{fig5} and \ref{fig6}, which show the 
elemental abundances from the various model components, and the masses of the various 
components (defined below) versus ionization parameter.

In order to account for 
the diversity in the importance of radiative excitation between 
various ions and elements,  in the low ionization parameter log($\xi$)=1 component, 
which is responsible for most of the O VII emission, the oxygen abundance 
in the low column density component (component 4 in table \ref{table4}) is  smaller than the oxygen abundance in the 
high column component (component 1 in table \ref{table4}) by a factor 2.5.   If not, the 
recombination emission into O VII would be stronger than observed. Iron abundances
are  similar between the low and high column density components, and are highest ($\sim$10)
at low ionization parameter (green in figure \ref{fig6}) and lowest ($\sim$1) at high 
ionization parameter log($\xi$)=2.6.  The normalizations and masses of the low 
column density components (components 4 -- 6) is generally lower than for the 
high column components (components 1 -- 3).  The normalization for component 5 is 
consistent with zero, though we include it for completeness.  Component 6 is needed 
to provide photoexcitation-dominated lines of Fe XIX -- XXI.

None of the abundances is significantly 
subsolar, except for O at log($\xi$)=2.6 and column=3 $\times 10^{22}$ cm$^{-2}$ 
(component 6). This is necessary to avoid over-producing O VIII RRCs.
It is worth noting that X-ray observations are not generally capable of determining 
abundances relative to H or He, but rather only abundances relative to other 
abundant metal elements such as O or Fe.  Thus, an apparent underabundance 
of O, for example, is equivalent to an overabundance of other elements such as Si, S, and 
Fe, relative to O.

The highest ionization parameter component is responsible 
for the Si and Mg emission.    Although its mass is dominated by the 
 3$\times 10^{23}$ cm $^{-2}$ component (component 3 in table \ref{table4}), 
the abundances of these 
elements are lower than for lower Z elements, so the optical depths in the He-like 
resonance lines are small enough that radiative excitation dominates the He-like 
lines.  It is notable that the differences 
in the importance of radiative excitation between the various He-like ions, and also Fe XVII, 
can essentially all be accounted for by the effects of varying amounts of resonance 
line optical depth which is produced solely as a result of ionization and elemental 
abundance effects.  This is consistent with the hypothesis suggested by \cite{Ogle03}.
An exception is for the lines from Ne IX, for which a large ion column 
density is needed in order to suppress the radiative excitation.
Models with column less than 3 $\times 10^{22}$ cm$^{-2}$ produce G$\leq$1.6, while the 
observations give G$\simeq$ 4$^{+6}_{-2}$. Our model produces the Ne IX 
$f$ line with  approximately half the the observed strength, while producing 
$r$ and $i$ lines which are close to the observed strengths.
The Ne IX RRC is also too weak in our models.  We have found, through 
experimentation, that an additional component with pure Ne and a 
column of 3 $\times 10^{24}$ cm$^{-3}$, and this is able to provide the 
observed Ne recombination features, but we have not included this component in the 
fits in figures \ref{fig1a} -- \ref{fig1d}  or table \ref{table4}.

The  table models are calculated assuming that the 
emitting gas fills a spherical shell defined by an inner radius 
$R_i=\left(\frac{L}{n\xi}\right)^{1/2}$ and outer radius $R_o$ defined by 
$\int_{R_i}^{R_o}n(R)dR=N$.
Table \ref{table4} provides the normalizations $\kappa_i$ of the table models 
as derived from the {\sc xspec} fits.  This normalization can be interpreted
in terms of the properties of the source as follows: $\kappa=f \frac{L_{38}}{D_{kpc}^2}$
where $L_{38}$ is the source ionizing 
luminosity integrated over the 1 - 1000 Ry energy range in units of 10$^{38}$ erg s$^{-1}$,
and $D_{kpc}$ is the distance to the source in kpc.  The normalization also includes 
a filling factor $f$ which accounts for the possibility that the gas does not 
fill the solid angle, and 
 $f \leq 1$.   We assume the density $n$ scales as $n=n_i(R/R_i)^{-2}$.
With this assumption the ionization parameter is independent of position 
for a given component and the total amount of gas in 
each emission component $j$, is:
\begin{equation}
M_j=4\pi f_j m_H n_iR_i^2(R_o-R_i)
\end{equation}
\noindent 
Similarly, the emission measure for each component can be written:

\begin{equation}
{\rm EM}_j=4\pi f_i n_i^2 R_i^4 \left(\frac{1}{R_i}-\frac{1}{R_o}\right)
\label{eq1}
\end{equation}

\noindent Since $n_iR_i^2=L/\xi$ both the mass and the emission measure 
depend on the physical size of the emission region, but not directly on density.  
They also depend on the ionizing luminosity.
Table \ref{table4} includes values for $f_i$, $M_i$ and EM$_i$ 
derived using these expressions, assuming $L=10^{44}$ erg s$^{-1}$.  
The $Chandra$ HETG spectrum of NGC 1068 constrains the density to be $\leq 10^{10}$ cm$^{-3}$, 
from the strength of the O VII forbidden line.  In the results given in 
table \ref{table4} we have chosen the inner and outer radii of the emission 
region to reflect simple conclusions from our data:  the inner 
radius is $R_i=$1 pc, based on physical arguments  about the location of the obscuring 
torus \citep{Krol88}, and the outer radius is chosen to be $R_o$=200 pc in order to 
reflect the observed extent of the X-ray emission, 
and this is roughly consistent with the extraction region width.
The density is chosen such that the inner radius of the line emitting 
gas is at 1 pc for all three emission components.  
Total masses and emission measures are also given.  The total emission measure 
is log(EM)=66.3
This is greater than would be required to simply emit the observed lines
if the gas had an emissivity optimized for the maximum line emission, 
owing to the fact that the table models include the gas needed 
to create the column density responsible for shielding the photoionization dominated 
gas from the effects of photoexcitation. Furthermore, most of the mass and emission 
measure comes from hydrogen, while the $Chandra$ HETG spectrum constrains only 
metals.  Values for mass and emission measure are calculated using the elemental 
abundances given in table \ref{table4}.  

It is also apparent that the assumptions 
described so far are not fully self-consistent.  That is, if density $\propto R^{-2}$ 
and if $R_o >> R_i$ then the column density is $\simeq n_i R_i$, and an
inner radius of 1pc cannot produce an arbitrary column density and 
ionization parameter.  A self consistent model with density $\propto R^{-2}$ requires
different inner radii for the different ionization parameter and column 
density components, and results in a total mass which is greater than 
that given in table \ref{table4}  by approximately 
a factor of 12.  The emission measure is insensitive to these assumptions.
Our subsequent discussion is based on the simple scenario given in table \ref{table4}:
$R_i$=1 pc, $R_o$=200 pc.

%

Figures \ref{fig1a} --  \ref{fig1d} show general 
consistency between most of the lines 
and a single choice for outflow velocity.  This corresponds to a net 
redshift of 0.0023$^+_-$0.0002, or an outflow velocity of 
450$^+_-$50 km s$^{-1}$ relative
to the systemic redshift of 0.00383 from \cite{Blan97}.  There are apparent 
discrepancies between some model line centroids and observed 
features when this redshift is adopted. lines which 
show apparent shifts which differ significantly from this, however.
An example is Mg XII L$\beta$, which has an apparent central wavelength 
of 7.150 $\AA$, as shown in table \ref{table2}.  The laboratory wavelength 
of this line is 7.110 $\AA$ \cite{Drak71}, which would correspond 
to an outflow velocity of 1688 km s$^{-1}$.  This is greater than 
other Doppler shifts in the spectrum, and the difference cannot 
be easily explained from uncertainties in the rest  wavelength.
Other possible line identifications include K$\alpha$ lines from 
Si ions such as Si III or Si IV.  These would be associated with 
gas at lower ionization parameter than the majority of the other 
lines in the spectrum and we have not attempted to include them in 
our model.

One result of global model fits is revealing the presence of features 
which are not obviously apparent from examination or Gaussian fits to the 
spectrum.  These are features, primarily from Fe L shell ions,
which are in the model, but which are 
blended or are not sufficiently statistically significant in 
the spectrum to require their inclusion in Gaussian fits.  
Therefore, these lines are not identified in table \ref{table2}.
Some are indicative of recombination vs. radiative excitation, 
in a sense similar to the Fe XVII lines discussed above.
Examples can be seen from examination of figures \ref{fig1a} -- \ref{fig1d}:
RRCs from Fe XVII and XVIII at 9.15 and 8.55 $\AA$, respectively, and 
Fe Lines at 10.55 (Fe XVII 2p-5d 10.523 $\AA$ in the lab), 
10.62, 10.85 (Fe XIX 2p-4d 10.6323,10.8267 $\AA$),
11.05 (Fe XVII  2s-4p 11.043 $\AA$), 
11.18, 11.28 (Fe XVII 2p-5d, 11.133, 11.253 $\AA$)
11.45 (Fe XVIII  2p-5s, 11.42$\AA$)
12.15 (Fe XVII  2p-4d 12.12 $\AA$), 
12.95 (Fe XIX 2s-3p, 12.92 $\AA$)
13.49, 13.53  (Fe XIX 2p-3d, 13.46,13.5249 $\AA$) 
14.3 (Fe XVIII 2p-3d, 14.208 $\AA$) 
15.88 (Fe XVIII 2p-3s, 15.83 $\AA$).

\section{Discussion}
\label{discussion}

It is possible to use the observed spatial extent of the X-ray emission to 
make further inferences about the X-ray emitting gas.  For NGC 1068, 
the distance is such that 1 arcsec = 72 pc \citep{Blan97}.  
The brightness profile in the zero order image is such that a significant 
amount of flux is coming from outside an extraction region  which is half the 
standard size, i.e. 2.4 arcsec.  Thus, 
an approximate, but convenient, size scale for the extent of the observed 
X-rays is $\simeq$ 2.4 arcsec $\simeq$ 200 pc.  If so, we can compare with 
the sizes derived from table \ref{table4}.   These estimates are based on an  assumed 
ionizing luminosity for the nucleus of 10$^{44}$ erg s$^{-1}$; this is uncertain 
by a factor of a few, but is bounded by $L_{bol}=4 \times 10^{44}$ erg s$^{-1}$ \citep{Blan97}.


The mass of material we infer based on the $Chandra$ HETG X-ray  spectrum is 
$\simeq 3.7 \times 10^5 M_\odot$, 
though this depends on the assumptions about the inner and outer radii of 
the emission region and on the assumption that density scales with $R^{-2}$.
This can be compared with typical masses for the entire narrow line region, 
which are likely to be $\sim 10^6 M_\odot$, although gas with ionization 
parameters in the range log($\xi$) 1 -- 3 will not radiate efficiently in 
typical optical/UV narrow lines.  If this gas flows uniformly 
over a distance $\simeq$ 200 pc at the speed we derive, then the 
flow timescale is 4.4 $\times 10^5$ yr, and the mean mass loss rate 
is 0.3 $M_\odot$ yr$^{-1}$.  This is considerably greater than the 
mass flux needed to power the nucleus, which is 
$\simeq$1.7 $\times 10^{-2} M_\odot$ yr$^{-1}$ $(\eta/0.1)^{-1}$ L/($10^{44}$ erg s$^{-1}$)
where $\eta$ is the efficiency of conversion of accreted mass into 
continuum luminosity.  

The filling factors in table \ref{table4} can be interpreted formally 
as the fraction of the total available volume in the spherical shell 
bounded by $R_i$ and $R_o$ which is filled with line emitting gas.  
The angular distribution of this gas is not constrained, so this can be 
interpreted as a conical region with fractional solid angle given by $f$ 
or as a spherically symmetric distribution in which the fractional 
covering of each component is given by $f$ when averaged over solid angle.
In either case it is clear that the total covering fraction of the 
reprocessing gas is at most a few percent.  This suggests 
that, if NGC 1068 were viewed from an arbitrarily chosen angle, and 
if the line of sight to the central continuum source were not blocked 
by the Compton thick torus, then the probability of observing a warm 
absorber similar to those seen in many Seyfert 1 galaxies would be small.
If the fractional solid angle of the obscuring torus as seen from 
the nucleus is $\sim$50$\%$ of the 
total, then the probability of seeing a warm absorber would be $\sim$10$\%$.  
This estimate is inversely proportional to the value we have assumed 
for the ionizing luminosity of the nucleus in NGC 1068, which is $10^{44}$ 
erg s$^{-1}$.  A smaller value for this quantity, which is possible
depending on the true spectral shape, would increase the inferred probability 
of seeing a warm absorber toward a value $\sim$50$\%$. 

Our model fits to the NGC 1068 HETG spectrum allow a simple 
phenomenological test of the scenario which has been widely used to 
interpret the spectra of Seyfert 1 objects.  That is, Seyfert 1 X-ray 
spectra, which show prominent blueshifted line absorption, have 
been analyzed assuming that the absorber lies solely along the line 
of sight and neglecting any effect of emission filling in 
the lines \citep{Mcke07}.  This can only be exactly correct if the 
absorber subtends negligible solid angle as seen from the 
central source.   This would in turn conflict with the apparent
presence of absorption in $\sim$half of known Seyfert 1 objects 
\citep{Reyn97}.  On the other hand, examination of the 
fits shown in figures \ref{fig1a} -- \ref{fig1d} and the results in table \ref{table2}
show a typical emission line flux for a strong line 
we measure is $\simeq$2 $\times 10^{-4}$ 
s$^{-1}$ cm$^{-2}$ for O VIII L$\alpha$.  This can be compared with 
the amount of energy absorbed in the same line from a Seyfert 1 
galaxy; for NGC 3783 this quantity is $\simeq$5 $\times 10^{-4}$ \citep{Kasp02}.
Correcting for the difference in distance, this would imply that, if 
the same emission line gas seen in NGC 1068 is also present in 
NGC 3783, then the ratio of line flux absorbed from the observed 
spectrum to emission is $\sim$8.  That is, the apparent flux
absorbed  in the O VIII L$\alpha$ line, and likely other lines as well,
is actually greater than observed by $\sim$ 10 -- 15 $\%$.  
The O VIII L$\alpha$ line in NGC 3783 shows signs of this with an apparent 
weak P-Cygni emission on the red edge of the trough, though this is not
apparent in most other lines in that spectrum.   Although O VIII L$\alpha$ 
is one of the strongest lines in the NGC 1068 spectrum, it is possible that
other lines in Seyfert 1 absorption spectra are more affected by filling 
in from emission.  Accurate fitting of these spectra, used to derive mass 
outflow rates and other quantities, would need to account for this process.

It is interesting to compare our results with those from 
UV and optical imaging and spectroscopy with the 
Faint Object Spectrograph (FOS) and  the Spave Telescope Imaging 
Spectrograph (STIS) on the Hubble Spate Telescope (HST). 
These properties have been discussed by 
\cite{Krae98,Cren00,Krae00,Cren00b,Krae00b}
The strongest UV emission lines observed by the HST instruments, eg. 
C IV $\lambda$1550, Si IV $\lambda$1398, and the He II and hydrogen lines,
comes from gas with a lower ionization parameter 
than the inferred from X-ray emitting gas. 
STIS provides spatially resolved kinematic information, which 
shows that the UV gas speed increases from the nucleus out 
to  130 pc, then the speed decreases at larger distances.
Also seen are coronal lines from S XII, Fe XIV Ne V, Ne IV.
These ions can exist at the same ionization parameter 
as the lowest ionization X-ray gas.  
The intensity as a function of distance from the nucleus 
shows apparent dilution as r$^-2$, and the emission region 
appears conical.   The continuum spectrum most closely resembles
a nuclear power law, implying that it is scattered light from 
the obscured nucleus.   The contribution from stars to the continuum 
is smaller than the non-thermal power law.

The gas observed by the HETG extends to $\sim$200 pc from the 
nucleus, while the obscuration has a much smaller size $\sim$1 pc.  
The gas which extends beyond the torus may originate near the torus, 
resembling a warm absorber flow in a type 1 object, and flowing 
ballistically to larger distances.  If so, 
the outflow speed would be expected to be constant or decrease with 
distance owing to gravitational forces, and the density would be 
subject to purely geometric dilution.  Alternatively, the 
X-ray and UV gas could entrain gas from narrow line clouds, 
or gas evaporated from the narrow line clouds, or from some 
other source.  Radiation pressure could play a role in 
energizing the flow.  The X-ray gas could play a role in the 
apparent deceleration of the UV gas seen at $\sim$100 pc.
There is little strong evidence for these latter 
scenarios from the HETG spectra, since we see no evidence 
for large changes in the ionization balance or speed of the gas 
in the spectra extracted from various width regions.  
Our results appear most nearly consistent with simple 
geometric dilution of the outflow, leading to constant ionization 
parameter with distance, and nearly constant outflow speed.

\section{Summary} 

The results presented in this paper can be summarized as follows:
(i) No single ionization parameter and excitation mechanism can 
fit to all the lines in the NGC 1068 X-ray spectrum obtained 
with the $Chandra$ HETG. 
The table model fits require three components at ionization 
parameters ranging from log($\xi$)=1 to  log($\xi$)=3.
We are able to adequately fit all the strong identified lines
in the spectrum with the exception of the lines from He-like 
Ne, which show a stronger recombination component than are 
produced by our models.  
(ii)  The abundances required by all the models are approximately solar 
\citep{Grev96} or slightly greater.  A notable exception is 
a large suppression of the abundance of oxygen at the highest ionization parameter.  Otherwise the O VIII RRCs would be stronger than observed. (iii) The 
masses and emission measures of gas are
greater than expected for optically thin emission, owing primarily  to the 
need for shielding to provide the recombination dominated gas 
seen in the lines of Mg and Si. (iv) The combined constraints on 
the ionization parameter and column density constrain the location 
of the emitting gas relative to the continuum source, while the line 
strengths constrain the amount of emitting gas.  Taken together, these 
require that the emitting gas have a volume filling factor 
less than unity; the values differ for the various components and range 
up to $\sim$0.01.  The mass flux through the 
region included in the HETG extraction region is approximately 
0.3 M$_\odot$ yr$^{-1}$ assuming ordered flow at the speed characterizing 
the line widths. (v) Limited experimentation with extracting spectra 
from various positions on the sky does not reveal a clear pattern 
which separates the various emitting components spatially.  
It appears that the emitting components coexist in the same 
physical region.

\appendix

\section{Appendix}

We make use of calculations 
using the {\sc xstar} \citep{Kal01} modeling package 
\footnote{http://heasarc.gsfc.nasa.gov/docs/software/xstar/xstar.html}.
{\sc xstar} is freely available 
and distributed as part of the {\sc heasoft} package. Models can be imported 
into {\sc xspec} and other fitting packages as tables, or via the 'analytic' model 
{\sc warmabs/photemis/scatemis}.  

Our models are based on the assumption that the most 
plausible energy source for the X-ray lines and RRCs 
observed from NGC 1068 is reprocessing of the continuum from the 
innermost regions of the AGN.  The continuum is presumably associated with the 
accretion disk and related structures close to the black hole, i.e.
within a few $R_G\simeq 3 \times 10^{11} M_6$ cm, where $M_6$ is the mass of the 
black hole in units of $10^6 M_\odot$.
The line and recombination emission we observe is likely formed 
at distances $ >10^6 R_G$, based on the geometry of the obscuring torus
as indicated by high spatial resolution IR imaging \citep{Jaff04}.
If it is assumed that the line emission is from gas where the heating,
excitation and ionization are dominated by the continuum from the black hole, 
and that there is a local time steady balance between these processes
and their inverses (radiative cooling, radiative decay and recombination),
then the  temperature, atomic level populations and ion fractions 
can be calculated.  
{\sc xstar} carries out such a calculation and also calculates the associated
X-ray emission and absorption. These emissivities and opacities 
are then applied to a simple one-dimensional solution of the equation of 
radiative transfer to derive the spectrum at a distant observer.  
Under  the simple assumption that the gas is optically thin the 
important physical quantities depend most sensitively on the ratio 
of the ionizing X-ray flux to the gas density.  We adopt the definition 
of this ionization parameter which is $\xi=4 \pi F_X/n$ where $F_X$ is the 
ionizing (energy) flux between  1 and 1000Ry and $n$ is the gas number 
density.   An important part of our results concerns the fact that the 
gas is not optically thin; this is key for explaining the varying amounts of 
radiative excitation seen in the line ratios.  This in turn means that the
column density, or the ion column densities, are also free parameters.
We implement the models by direct fitting, in which {\sc xstar}
models are used to generate synthetic spectra which are compared 
to the observations within the fitting program  {\sc xspec}.

{\sc xstar} explicitly calculates the populations of all atomic levels associated 
with emission or absorption of radiation; it does not rely on the traditional 
'nebular approximation' which assumes that every excitation decays only to ground.
Therefore it is straightforward to include radiative excitation, and this has 
been done.  Interactive, iterative calculation of full {\sc xstar} models within 
{\sc xspec} and simultaneous fitting to data is not practical owing 
to computational limitations.   

We point out, parenthetically, that the simplest way to use {\sc xstar} results within {\sc xspec} 
is to use the associated {\sc xspec} `analytic' models.
When called from {\sc xspec},  these read stored tables of ionic level populations as a function 
of ionization parameter, and then calculate the opacity and emissivity 
`on the fly'.  The physical quantity used the {\sc xspec} model fit 
is an emitted flux or a transmission coefficient as a function of energy for 
a photoionized slab of given column density, abundances and ionization parameter.
These are calculated from the opacity and emissivity by assuming these quantities are 
 uniform throughout.
Advantages of this procedure compared with the use of table models
(described below) include:
Ability to account for arbitrary element abundances, arbitrary spectral 
resolution,  and arbitrary turbulent broadening. 
Limitations include the fact that  it  uses a saved file of 
level populations calculated for a grid of optically thin models for a fixed 
choice of ionizing spectrum
rather than calculating the ionization balance self-consistently.
It implicitly assumes that the absorber has uniform ionization even if the user 
specifies a large column, 
and that all emission freely escape.  In this sense it is 
not self-consistent.  In fact, analytic models fail when applied to the analysis 
of the NGC 1068 spectrum because, for column densities large enough to suppress 
radiative excitation, as demanded by eg. the Fe XVII 17$\AA$ line, the associated 
RRCs for Fe XVII are assumed to escape freely and are greatly over predicted.
For this reason we will not discuss analytic models further in this paper.

In order to self-consistently include the effects of absorption 
of the incident continuum in models for photoionized emission from 
NGC 1068 it is necessary to use optically thick models.  These 
are calculated using {\sc xstar}, by performing multiple model runs 
and using varying ionization 
parameter and cloud column density.  The resulting emission spectra
and transmission functions  are stored as fits tables, binned in energy, 
using the table format prescribed by {\sc xspec}.  These can be 
read by {\sc xspec}, selected according to parameter values, and
interpolation between modeled parameter values is performed.
{\sc xspec} can fit the table models to the observed spectrum 
and thereby yield values for the cloud column density, ionization parameter,
and the normalization, which is described in more detail below.
Owing to the fact that the spectra are stored binned in energy the 
line width cannot be conveniently  treated as a free parameter, 
and therefore it is important 
that the energy grid spacing used in constructing the table 
be no greater than than the intrinsic instrumental resolution.   
It is also possible to treat the element abundances as free parameters, 
by assuming that the escaping line and RRC fluxes fluxes depend linearly 
on the abundances.  This ignores the coupling of the cloud ionization 
structure to abundance, via the temperature and the opacity of the model.
In our fits we use this approximation when deriving element abundances.  

Within a given optically thick model, {\sc xstar} calculates the effect 
of attenuation of the incident continuum using a simple single stream 
treatment of the radiative transfer.  This takes into account the 
varying ionization and excitation through the model.
The inclusion of radiative excitation 
necessitates a finer spatial grid than for continuum absorption alone, 
since the transfer must resolve the 
attenuation length of the photons in the resonance lines. 
Also, line opacity varies over a large range in a very narrow
energy range and it is not feasible to accurately sample the relevant energies for 
all of the many lines in a photoionized plasma.  We adopt a very simple 
treatment in which the line opacity is binned into continuum bins 
and used to calculate the attenuated flux.  This flux in continuum bins  is 
then used in the calculation of photoexcitation.   The continuum bin size 
is typically $\Delta\varepsilon/\varepsilon=1.4 \times 10^{-4}$ which corresponds 
to $\simeq 41$km s$^{-1}$ Doppler width.  This is comparable to the thermal 
Doppler width for, eg., oxygen at a temperature $\sim10^6$K.  Thus, we are not 
fully resolving most lines relevant to this study.  The tables used to fit to 
the $Chandra$ HETG spectrum of NGC 1068 include emission from both the illuminated 
and unilluminated cloud faces, corresponding to the assumption that the 
the line emitting material is arranged with approximate spherical symmetry
around the central continuum source.
They are calculated assuming an ionizing spectrum which is a single power law 
with (photon number) spectral index $\Gamma$=2 and constant density 10$^4$ cm$^{-3}$.  These results are not expected to depend sensitively on the latter 
two assumptions; in particular $\Gamma$ values in the range 1.7 -- 2.3 produce
very similar ionization balance distributions.

{\sc xstar} makes use of atomic data compiled from various sources 
and described by  \citep{Baut01}.   This has been extensively updated 
since that time, as described by \citet{Witt07,Witt09,Witt11,Palm02,Palm03,
Palm03b,Palm08a,Palm08b,Palm11,Palm12,Mend04,Garc09,Baut03,Baut04}
The results presented here which flow from the model calculations 
are dependent on the assumptions, computational implementation and 
atomic data used by the models.  The atomic data affects the line identifications
and the outflow speeds, via the line rest wavelengths.  It also affects the 
ionization balance, via the photoionization and recombination rates, and the 
line strengths.  The quantitative uncertainty associated with the atomic 
data and the resulting uncertainty in the synthetic spectrum are difficult 
to estimate; this is a topic of interest for many problems beyond this one.
We can point out that the emissivities of most of the lines from the H- and He-like
ions in our study depend on two processes: recombination and photoexcitation.
These depend in turn on photoionization cross sections and line oscillator strengths, 
respectively.  These are quantities associated with radiative processes in relatively 
simple ions and therefore are generally more reliable than collisional 
rate coefficients which are important for ions with partially filled L shells 
in our models and in coronal plasmas.  \cite{Fost10} have shown that uncertainties 
of, say, 20$\%$ on the collision strengths for O VII can result in a range 
of a factor of $\simeq$3 in the inferred temperature in a coronal plasma based on the 
$G$ ratio.  We do not anticipate similar sensitivity to atomic data uncertainties in
the NGC 1068 models, since the rates for recombination and photoexcitation 
do not have the strong non-linear dependence on temperature which is intrinsic 
to  collisional rates.   In particular, we do not consider the rates embodied in our models
to be sufficiently uncertain to change our results qualitatively, or to allow 
significantly better fits with fewer components or parameters.  A more likely 
shortcoming of our models is the neglect of some important physical mechanism.  
Examples include:  charge transfer; photo-excitation by some unseen source 
of radiation; non-ionization equilibrium effects; or inhomogeneities in 
elemental compositions associated with formation or destruction of dust.

\begin{deluxetable}{crrrrrrrrrrr}
\tabletypesize{\scriptsize}
\tablecaption{Observation log \label{table1}}
\tablewidth{0pt}
\tablehead{
\colhead{obsid} & \colhead{time} 
  & \colhead{exposure}   }
\startdata
332&2000-12-04 18:11:52&46290\\
9148&2008-12-05 08:23:41&80880\\
9149&2008-11-19 04:49:51&90190\\
9150&2008-11-27 04:55:36&41760\\
10815&2008-11-20 16:23:22&19380\\
10816&2008-11-18 01:18:39&16430\\
10817&2008-11-22 17:36:37&33180\\
10823&2008-11-25 18:21:16&35110\\
10829&2008-11-30 20:16:45&39070\\
10830&2008-12-03 15:08:36&43600\\
\enddata
\end{deluxetable}

\begin{deluxetable}{clllllllllllllll}
\tabletypesize{\scriptsize}
\tablecaption{Lines Found \label{table2}}
\tablewidth{0pt}
\tablehead{
\colhead{index} &\colhead{wavelength ($\AA$)} & \colhead{width (km/s)}   & \colhead{flux (cm$^{-2}$s$^{-1}$)}    & \colhead{lab  ($\AA$)}    & \colhead{ion}    & \colhead{lower level}    & \colhead{upper level}    & \colhead{v$_{off}$ (km/s)}   }
\startdata
   1& 1.780&$\leq$ 2400               &1.3$^{+0.1}_{-0.3} \times 10^{-5}$& 1.780&fe xxvi &1s$^1(^2$S)                   &1s$^0$2p$^1(^2$P)             &  1100$^{+   840}_{-    40}$ \\
   2& 1.855&$\leq$ 2000               &2.9$^{+0.2}_{-0.2} \times 10^{-5}$& 1.869&fe xxv  &1s$^2(^1$S)                   &1s$^1$2p$^1(^1$P)             &  3400$^{+    40}_{-   400}$ \\
   3& 1.945& 1700$^{+   17}_{-   17}$ &5.6$^{+0.2}_{-0.3} \times 10^{-5}$& 1.941&fe i    &K$\alpha$                     &                              &   500$^{+    40}_{-   800}$ \\
   4& 2.550&$\leq$ 2500               &2.7$^{+0.1}_{-1.6} \times 10^{-6}$& 2.549&ca xx   &1s$^1(^2$S)                   &1s$^0$3p$^1(^2$P)             &  1100$^{+  2400}_{-    40}$ \\
   5& 3.030& 1900$^{+  550}_{-  440}$ &3.9$^{+0.1}_{-1.6} \times 10^{-6}$& 3.020&ca xx   &1s$^1(^2$S)                   &1s$^0$2p$^1(^2$P)             &   150$^{+    40}_{-  1000}$ \\
   6& 3.175&$\leq$ 3600               &4.7$^{+0.4}_{-0.4} \times 10^{-6}$& 3.150&ar xviii&1s$^1(^2$S)                   &1s$^0$3p$^1(^2$P)             & -                           \\
   7& 3.195&$\leq$ 2600               &2.3$^{+0.4}_{-0.4} \times 10^{-6}$& 3.180&ca xix  &1s$^2(^1$S)                   &1s$^1$2p$^1(^1$P)             & -                           \\
   8& 3.195&$\leq$ 2600               &2.3$^{+0.4}_{-0.4} \times 10^{-6}$& 3.190&ca xix  &1s$^2(^1$S)                   &1s$^1$2p$^1(^3$P)             &   670$^{+  3300}_{-    40}$ \\
   9& 3.215&$\leq$ 2000               &2.7$^{+0.4}_{-0.4} \times 10^{-6}$& 3.210&ca xix  &1s$^2(^1$S)                   &1s$^1$2s$^1(^3$S)             &   670$^{+  5200}_{-    40}$ \\
  10& 3.380&$\leq$ 2200               &5.5$^{+0.4}_{-0.5} \times 10^{-6}$& 3.365&ar xvii &1s$^2(^1$S)                   &1s$^1$3p$^1(^1$P)             & -                           \\
  11& 3.735&$\leq$ 2200               &4.0$^{+0.1}_{-1.4} \times 10^{-6}$& 3.739&ar xviii&1s$^1(^2$S)                   &1s$^0$2p$^1(^2$P)             &  1500$^{+   840}_{-   400}$ \\
  12& 3.945&$\leq$ 2000               &5.1$^{+0.2}_{-1.5} \times 10^{-6}$& 3.950&ar xvii &1s$^2(^1$S)                   &1s$^1$2p$^1(^1$P)             &  1500$^{+  1500}_{-    40}$ \\
  13& 4.000&$\leq$ 2600               &5.2$^{+0.5}_{-0.4} \times 10^{-6}$& 3.990&ar xvii &1s$^2(^1$S)                   &1s$^1$2p$^1(^3$P)             &   380$^{+  1300}_{-    40}$ \\
  14& 4.000&$\leq$ 2600               &5.2$^{+0.5}_{-0.4} \times 10^{-6}$& 3.990&s xvi   &1s$^1(^2$S)                   &1s$^0$3p$^1(^2$P)             &   380$^{+  1300}_{-    40}$ \\
  15& 4.000&$\leq$ 2600               &5.2$^{+0.5}_{-0.4} \times 10^{-6}$& 4.010&ar xvii &1s2.1S                        &1s1.2s1.3S                    &  1900$^{+  1300}_{-    40}$ \\
  16& 4.755&$\leq$ 2100               &7.6$^{+0.4}_{-0.9} \times 10^{-6}$& 4.730&s xvi   &1s1.2S                        &1s0.2p1.2P                    & -                           \\
  17& 5.040&$\leq$ 2100               &1.4$^{+0.1}_{-0.1} \times 10^{-5}$& 5.040&s xv    &1s$^2(^1$S)                   &1s$^1$2p$^1(^1$P)             &  1100$^{+  1500}_{-    40}$ \\
  18& 5.100&$\leq$ 2000               &8.6$^{+0.8}_{-0.7} \times 10^{-6}$& 5.070&s xv    &1s$^2(^1$S)                   &1s$^1$2p$^1(^3$P)             & -                           \\
  19& 5.100&$\leq$ 2000               &8.6$^{+0.8}_{-0.7} \times 10^{-6}$& 5.084&si xiii &rrc                           &                              &   180$^{+  2400}_{-   160}$ \\
  20& 5.225&$\leq$ 2000               &8.6$^{+0.7}_{-0.6} \times 10^{-6}$& 5.212&si xiv  &1s$^1(^2$S)                   &1s$^0$3p$^1(^2$P)             &   390$^{+    40}_{-   320}$ \\
  21& 5.690& 1100$^{+  120}_{-  130}$ &7.0$^{+0.3}_{-1.1} \times 10^{-6}$& 5.680&si xiii &1s$^2(^1$S)                   &1s$^1$3p$^1(^1$P)             &   600$^{+   160}_{-    40}$ \\
  22& 6.195& 1300$^{+  150}_{-   27}$ &2.2$^{+0.0}_{-0.2} \times 10^{-5}$& 6.180&si xiv  &1s$^1(^2$S)                   &1s$^0$2p$^1(^2$P)             &   410$^{+   280}_{-   120}$ \\
  23& 6.660& 1500$^{+  120}_{-   47}$ &3.1$^{+0.0}_{-0.2} \times 10^{-5}$& 6.640&si xiii &1s$^2(^1$S)                   &1s$^1$2p$^1(^1$P)             &   230$^{+   480}_{-    40}$ \\
  24& 6.750& 1700$^{+ 1300}_{- 2300}$ &2.4$^{+0.1}_{-0.1} \times 10^{-5}$& 6.690&si xiii &1s$^2(^1$S)                   &1s$^1$2p$^1(^3$P)             & -                           \\
  25& 6.750& 1700$^{+ 1300}_{- 2300}$ &2.4$^{+0.1}_{-0.1} \times 10^{-5}$& 6.744&si xiii &1s$^2(^1$S)                   &1s$^1$2s$^1(^3$S)             &   870$^{+   120}_{-   120}$ \\
  26& 6.870&$\leq$ 2000               &7.4$^{+0.4}_{-0.3} \times 10^{-6}$& 6.860&si x    &K$\alpha$                     &                              &   700$^{+    40}_{-   120}$ \\
  27& 6.995&$\leq$ 3600               &7.3$^{+0.1}_{-2.2} \times 10^{-6}$& 6.930&si ix   &K$\alpha$                     &                              & -                           \\
  28& 7.005&  450                     &2.9$^{+2.5}_{-1.6} \times 10^{-7}$& 7.001&si viii &K$\alpha$                     &                              &   970$^{+  5600}_{-    40}$ \\
  29& 7.005&  450                     &2.9$^{+2.5}_{-1.6} \times 10^{-7}$& 7.040&mg xi   &rrc                           &                              &  2600$^{+  5600}_{-    40}$ \\
  30& 7.150&$\leq$ 2700               &2.2$^{+0.1}_{-0.1} \times 10^{-5}$& 7.110&mg xii  &1s$^1(^2$S)                   &1s$^0$3p$^1(^2$P)             & -                           \\
  31& 7.320&$\leq$ 2000               &6.1$^{+0.4}_{-0.3} \times 10^{-6}$& 7.310&mg xi   &1s$^2(^1$S)                   &1s$^1$5p$^1(^1$P)             &   740$^{+   640}_{-    40}$ \\
  32& 7.485& 1400$^{+   14}_{-   14}$ &8.1$^{+0.2}_{-1.3} \times 10^{-6}$& 7.470&mg xi   &1s$^2(^1$S)                   &1s$^1$4p$^1(^1$P)             &   530$^{+    40}_{-   240}$ \\
  33& 7.770&$\leq$ 2000               &9.6$^{+0.4}_{-0.5} \times 10^{-6}$& 7.758&al xii  &1s$^2(^1$S)                   &1s$^1$3p$^1(^1$P)             &   640$^{+  4400}_{-    40}$ \\
  34& 7.860& 1200$^{+  200}_{-   12}$ &1.2$^{+0.1}_{-0.0} \times 10^{-5}$& 7.850&mg xi   &1s$^2(^1$S)                   &1s$^1$3p$^1(^1$P)             &   750$^{+  1200}_{-    40}$ \\
  35& 8.435& 1400$^{+   96}_{-   43}$ &3.3$^{+0.0}_{-0.3} \times 10^{-5}$& 8.420&mg xii  &1s$^1(^2$S)                   &1s$^0$2p$^1(^2$P)             &   600$^{+   360}_{-    40}$ \\
  36& 9.190& 1500$^{+  230}_{-   30}$ &4.2$^{+0.0}_{-0.4} \times 10^{-5}$& 9.170&mg xi   &1s$^2(^1$S)                   &1s$^1$2p$^1(^1$P)             &   480$^{+   200}_{-   120}$ \\
  37& 9.340&$\leq$ 2500               &3.5$^{+0.1}_{-0.1} \times 10^{-5}$& 9.230&mg xi   &1s$^2(^1$S)                   &1s$^1$2p$^1(^3$P)             & -                           \\
  38& 9.340&$\leq$ 2500               &3.5$^{+0.1}_{-0.1} \times 10^{-5}$& 9.320&mg xi   &1s$^2(^1$S)                   &1s$^1$2s$^1(^3$S)             &   490$^{+    40}_{-   760}$ \\
  39& 9.465&$\leq$ 2000               &1.9$^{+0.0}_{-0.3} \times 10^{-5}$& 9.475&fe xxi  &2p$^2(^3$P$_0$)               &2p$^1$4d$^1(^3$D$_1$)         &  1500$^{+  1900}_{-    80}$ \\
  40& 9.720&$\leq$ 2000               &1.9$^{+0.1}_{-0.1} \times 10^{-5}$& 9.679&fe xix  &2p$^4(^1$D$_2$)               &2p$^3$5d$^1(^3$D$_3\#3$)      & -                           \\
  41& 9.720&$\leq$ 2000               &1.9$^{+0.1}_{-0.1} \times 10^{-5}$& 9.800&fe xvii &rrc                           &                              &  3600$^{+   160}_{-    79}$ \\
  42&10.045&$\leq$ 2400               &2.5$^{+0.0}_{-0.7} \times 10^{-5}$&10.014&fe xix  &2p$^4(^1$D$_2$)               &2p$^3$5d$^1(^3$D$_3$)         &   230$^{+    40}_{-   400}$ \\
  43&10.260&$\leq$ 2000               &3.9$^{+0.1}_{-0.2} \times 10^{-5}$&10.240&ne x    &1s$^1(^2$S)                   &1s$^0$3p$^1(^2$P)             &   550$^{+  1600}_{-    40}$ \\
  44&10.365& 1600$^{+   95}_{-  120}$ &3.4$^{+0.0}_{-0.7} \times 10^{-5}$&10.388&ne ix   &rrc                           &                              &  1800$^{+   320}_{-    40}$ \\
  45&10.670&$\leq$ 2200               &3.6$^{+0.2}_{-0.1} \times 10^{-5}$&10.641&fe xix  &2p$^4(^3$P$_2$)               &2p$^3$4d$^1(^3$P$_2$)         &   320$^{+    40}_{-  1300}$ \\
  46&10.670&$\leq$ 2200               &3.6$^{+0.2}_{-0.1} \times 10^{-5}$&10.660&fe xvii &                              &                              &   860$^{+    40}_{-  1300}$ \\
  47&10.830&$\leq$ 2000               &2.3$^{+0.1}_{-0.1} \times 10^{-5}$&10.770&fe xvii &2p$^6(^1$S$_0$)               &2p$^5$6d$^1(^1$P$_1$)         & -                           \\
  48&10.830&$\leq$ 2000               &2.3$^{+0.1}_{-0.1} \times 10^{-5}$&10.820&fe xix  &                              &                              &   860$^{+   240}_{-    40}$ \\
  49&11.025& 1300$^{+  300}_{-   13}$ &3.5$^{+0.1}_{-0.4} \times 10^{-5}$&11.000&ne ix   &1s$^2(^1$S)                   &1s$^1$4p$^1(^1$P)             &   460$^{+   160}_{-    80}$ \\
  50&11.555& 1400$^{+   40}_{-   70}$ &4.0$^{+0.0}_{-0.6} \times 10^{-5}$&11.500&fe xviii&                              &                              & -                           \\
  51&11.555& 1400$^{+   40}_{-   70}$ &4.0$^{+0.0}_{-0.6} \times 10^{-5}$&11.547&ne ix   &1s$^2(^1$S)                   &1s$^1$3p$^1(^1$P)             &   930$^{+   280}_{-    40}$ \\
  52&11.800&$\leq$ 2000               &4.2$^{+0.1}_{-0.2} \times 10^{-5}$&11.762&fe xxi  &2p$^2(^3$P$_1$)               &2s$^1$2p$^2$3d$^1(^3$P$_2$)   &   170$^{+   520}_{-    80}$ \\
  53&12.165& 1400$^{+  200}_{-   14}$ &9.5$^{+0.1}_{-0.8} \times 10^{-5}$&12.100&ne x    &1s$^1(^2$S)                   &1s$^0$2p$^1(^2$P)             & -                           \\
  54&12.300&$\leq$ 2400               &5.7$^{+0.1}_{-1.6} \times 10^{-5}$&12.264&fe xvii &2p$^6(^1$S$_0$)               &2p$^5$4d$^1(^3$D$_1$)         &   260$^{+    40}_{-   280}$ \\
  55&12.840&$\leq$ 2000               &5.5$^{+0.2}_{-0.3} \times 10^{-5}$&12.800&fe xx   &                              &                              &   200$^{+  1300}_{-    40}$ \\
  56&12.840&$\leq$ 2000               &5.5$^{+0.2}_{-0.3} \times 10^{-5}$&12.812&fe xviii&2p$^5(^2$P$_{3/2}$)           &2s$^1$2p$^5$3p$^1(^2$D$_5$)   &   480$^{+  1300}_{-    40}$ \\
  57&13.509&$\leq$ 3600               &3.2$^{+0.3}_{-0.5} \times 10^{-5}$&13.447&ne ix   &1s$^2(^1$S)                   &1s$^1$2p$^1(^1$P)             & -                           \\
  58&13.509&$\leq$ 3600               &3.2$^{+0.3}_{-0.5} \times 10^{-5}$&13.500&ne ix   &1s$^2(^1$S)                   &1s$^1$2p$^1(^3$P)             &   940$^{+  6700}_{-    40}$ \\
  59&13.725&$\leq$ 3000               &1.4$^{+0.0}_{-0.1} \times 10^{-4}$&13.700&ne ix   &1s$^2(^1$S)                   &1s$^1$2s$^1(^3$S)             &   590$^{+   360}_{-    40}$ \\
  60&14.215&$\leq$ 2000               &1.1$^{+0.0}_{-0.2} \times 10^{-4}$&14.206&fe xviii&2p$^5(^2$P$_{3/2}$)           &2p$^5$3d$^1(^2$D$_{5/2}\#$2)  &   950$^{+   960}_{-    40}$ \\
  61&14.215&$\leq$ 2000               &1.1$^{+0.0}_{-0.2} \times 10^{-4}$&14.250&o viii  &rrc                           &                              &  1900$^{+   960}_{-    40}$ \\
  62&14.415&$\leq$ 2100               &7.2$^{+0.1}_{-2.2} \times 10^{-5}$&14.394&fe xviii&2p$^5(^2$P$_{3/2}$)           &2p$^4$3d$^1(^2$D$_{5/2}$)     &   710$^{+    40}_{-   960}$ \\
  63&14.580&$\leq$ 2000               &5.5$^{+0.4}_{-0.3} \times 10^{-5}$&14.500&fe xx   &2p$^3(^2$P$_{1/2}$)           &2p$^2$3s$^1(^4$P$_{1/2}$)     & -                           \\
  64&14.835& 1600$^{+  410}_{-  100}$ &3.5$^{+0.1}_{-0.8} \times 10^{-5}$&14.816&fe xix  &2p$^4(^3$P$_1$)               &2p$^3$3s$^1(^1$D$_2$)         &   740$^{+   920}_{-    40}$ \\
  65&14.835& 1600$^{+  410}_{-  100}$ &3.5$^{+0.1}_{-0.8} \times 10^{-5}$&14.832&o viii  &1s1.2S                        &1s0.5p1.2P                    &  1100$^{+   920}_{-    40}$ \\
  66&15.040&$\leq$ 2700               &1.2$^{+0.1}_{-0.0} \times 10^{-4}$&15.015&fe xvii &2p$^6(^1$S$_0$)               &2p$^5$3d$^1$                  &   640$^{+  1600}_{-    40}$ \\
  67&15.295&$\leq$ 3600               &8.5$^{+0.4}_{-0.3} \times 10^{-5}$&15.188&o viii  &1s$^1(^2$S)                   &1s$^0$4p$^1(^2$P)             & -                           \\
  68&15.295&$\leq$ 3600               &8.5$^{+0.4}_{-0.3} \times 10^{-5}$&15.262&fe xvii &2p$^6(^1$S$_0$)               &2p$^5$3d$^1(^3$P$_1$)         &   490$^{+    40}_{-   990}$ \\
  69&15.295&$\leq$ 3600               &8.5$^{+0.4}_{-0.3} \times 10^{-5}$&15.418&fe xvii &2p$^6(^1$S$_0$)               &2p$^5$3d$^1(^3$P$_2$)         &  3500$^{+    39}_{-   980}$ \\
  70&15.670&$\leq$ 2800               &2.5$^{+0.2}_{-0.2} \times 10^{-5}$&15.627&fe xviii&2p$^5(^2$P$_{3/2}$)           &2p$^4$3s$^1(^2$P$_{5/2}$)     &   310$^{+   200}_{-   200}$ \\
  71&16.040& 1500$^{+  250}_{-   62}$ &8.4$^{+0.2}_{-1.0} \times 10^{-5}$&16.007&fe xviii&2p$^5(^2$P$_{3/2}$)           &2p$^4$3s$^1(^2$P$_{3/2}$)     &   520$^{+   400}_{-   120}$ \\
  72&16.785& 1700$^{+  500}_{-   17}$ &1.8$^{+0.0}_{-0.3} \times 10^{-4}$&16.777&fe xvii &2p$^6(^1$S$_0$)               &2p$^5(^3$S$_1$)               &   990$^{+    40}_{-   160}$ \\
  73&16.785& 1700$^{+  500}_{-   17}$ &1.8$^{+0.0}_{-0.3} \times 10^{-4}$&16.777&o vii   &rrc                           &                              &   990$^{+    40}_{-   160}$ \\
  74&17.135&$\leq$ 2000               &1.6$^{+0.1}_{-0.1} \times 10^{-4}$&17.050&fe xvii &2p$^6(^1$S$_0$)               &2p$^5(^3$S$_1$)               & -                           \\
  75&17.420& 1100$^{+  300}_{-   58}$ &3.1$^{+0.2}_{-0.4} \times 10^{-5}$&17.396&o vii   &1s$^2(^1$S)                   &1s$^1$5p$^1(^1$P)             &   720$^{+    40}_{-   200}$ \\
  76&17.785& 1400$^{+  200}_{-  160}$ &5.2$^{+0.3}_{-0.5} \times 10^{-5}$&17.768&o vii   &1s$^2(^1$S)                   &1s$^1$4p$^1(^1$P)             &   850$^{+   440}_{-    40}$ \\
  77&18.635& 1700$^{+  470}_{-   34}$ &1.2$^{+0.0}_{-0.2} \times 10^{-4}$&18.627&o vii   &1s$^2(^1$S)                   &1s$^1$3p$^1(^1$P)             &  1000$^{+    40}_{-   280}$ \\
  78&19.025& 1200$^{+   56}_{-   59}$ &3.1$^{+0.1}_{-0.2} \times 10^{-4}$&18.968&o viii  &1s$^1(^2$S)                   &1s$^0$2p$^1(^2$P)             &   240$^{+    80}_{-   160}$ \\
  79&20.970& 1500$^{+  360}_{-  190}$ &9.0$^{+0.8}_{-1.4} \times 10^{-5}$&20.910&n vii   &1s$^1(^2$S)                   &1s$^0$3p$^1(^2$P)             &   280$^{+    40}_{-   600}$ \\
  80&21.651& 1300$^{+  230}_{-   38}$ &2.4$^{+0.1}_{-0.2} \times 10^{-4}$&21.602&o vii   &1s$^2(^1$S)                   &1s$^1$2p$^1(^1$P)             &   460$^{+   440}_{-    40}$ \\
  81&21.855&$\leq$ 2900               &1.7$^{+0.1}_{-0.6} \times 10^{-4}$&21.804&o vii   &1s$^2(^1$S)                   &1s$^1$2p$^1(^3$P)             &   440$^{+    40}_{-  1000}$ \\
  82&22.145&  560$^{+   27}_{-   11}$ &5.8$^{+0.5}_{-0.2} \times 10^{-4}$&22.110&o vii   &1s$^2(^1$S)                   &1s$^1$2s$^1(^3$S)             &   660$^{+    80}_{-    40}$ \\
  83&23.625&$\leq$ 3600               &6.5$^{+0.4}_{-2.1} \times 10^{-5}$&23.440&o i     &2p$^4(^1$D$_2$)               &1s$^1$2s$^2$2p$^5(^1$P$_1$)   & -                           \\
  84&23.835&$\leq$ 2600               &8.6$^{+0.6}_{-2.2} \times 10^{-5}$&23.771&n vi    &1s$^2(^1$S)                   &1s$^1$4p$^1(^1$P)             &   330$^{+   160}_{-   720}$ \\
  85&24.840& 1300$^{+  210}_{-   38}$ &3.1$^{+0.1}_{-0.3} \times 10^{-4}$&24.781&n vii   &1s$^1(^2$S)                   &1s$^0$2p$^1(^2$P)             &   420$^{+   200}_{-    40}$ \\
  86&24.840& 1300$^{+  210}_{-   38}$ &3.1$^{+0.1}_{-0.3} \times 10^{-4}$&25.164&ar xv   &2s$^2(^1$S$_0$)               &2s$^1$3p$^1(^3$P$_1$)         &  5000$^{+   200}_{-    39}$ \\
\enddata
\end{deluxetable}

\begin{table}
\begin{center}
\caption{Model for Fe K region Parameters \label{table5}}
\begin{tabular}{llllllllllll}
\tableline
\tableline
component&log($\xi$)&norm&width (km/s) &velocity (km/s)\\
photemis&-3&2.47$^{+0.82}_{-0.61} \times 10^8$&1.28$^{+0.25}_{-0.23}\times 10^3$&2.41$^{+0.51}_{-0.67} \times 10^2$ km s$^{-1}$\\
scatemis&3&33.1$^{+8.7}_{-9.7}$&1.28$^{+0.25}_{-0.23}\times 10^3$&2.41$^{+0.51}_{-0.67} \times 10^2$ km s$^{-1}$\\
\tableline
\end{tabular}


\end{center}
\end{table}

\begin{deluxetable}{clllllllllllllll}
\tabletypesize{\scriptsize}
\rotate
\tablecaption{Fitting Results:  Table Models \label{table4}}
\tablewidth{0pt}
\tablehead{
\colhead{Component} & \colhead{1}   & \colhead{2}    & \colhead{3}    & \colhead{4}    & \colhead{5}    & \colhead{6}    & \colhead{Total}   }
\tabletypesize{\scriptsize}
\startdata
log(N) & 23.5   & 23.5   & 23.5   & 22.5   & 22.5   & 22.5   & \\
log($\xi$) & 1   & 1.8   & 2.6   & 1   & 1.8   & 2.6   & \\
N \tablenotemark{a}& 0.99 $\pm{ 0.19 }$& -   & -   & -   & -   & -   & \\
O \tablenotemark{a}& 2.48 $\pm{ 0.3 }$& 0.86 $\pm{ 0.6 }$& $\leq$0.09& 1.1 $\pm{ 1.5 }$&$\leq$10& $\leq$0.1& \\
Ne \tablenotemark{a}& 1.91 $\pm{ 0.15 }$& -   & -   & -   & -   & -   & \\
Mg \tablenotemark{a}& 1.06 $\pm{ 0.2 }$& -   & -   & -   & -   & -   & \\
Si \tablenotemark{a}& 1.49 $\pm{ 0.1 }$& -   & -   & -   & -   & -   & \\
S \tablenotemark{a}& 1.41 $\pm{ 0.3 }$& -   & -   & -   & -   & -   & \\
Ar \tablenotemark{a}& 1.88 $\pm{ 0.5 }$& -   & -   & -   & -   & -   & \\
Ca \tablenotemark{a}& 2.42 $\pm{ 0.9 }$& -   & -   & -   & -   & -   & \\
Fe \tablenotemark{a}& $\leq$100& 2.7 $\pm{ 0.5 }$& 1 $\pm{ 0.15 }$& 8 $\pm{ 1 }$&$\leq$100& 1.88 $\pm{ 1.68 }$& \\
$\kappa$ ($\times 10^{-6}$) & 16.4 $\pm{ 3 }$& 22. $\pm{ 3 }$& 83. $\pm{ 6 }$& 6.7 $\pm{ 1 }$& $\leq$0.1& 3.9 $\pm{ 0.4 }$& \\
f ($\times 10^{-4}$) & 34.0 $\pm{ 6.2 }$& 45.6 $\pm{ 6.4 }$& 172. $\pm{ 12.4 }$& 13.9 $\pm{ 2.1 }$& $\leq$0.1& 8.0 $\pm{ 0.8 }$& \\
mass($\times 10^{4} M_\odot$)\tablenotemark{b} & 21.3 $\pm{ 3.9 }$& 4.5 $\pm{ 0.6 }$& 2.7 $\pm{ 0.2 }$& 8.7 $\pm{ 1.3 }$& $\leq$0.02 & 0.13 $\pm{ 0.01 }$&37.3  $\pm{6.0}$ \\
log(EM (cm$^{-3}$)) & 66.2 $\pm{ 0.07 }$& 64.7 $\pm{ 0.06}$& 63.7 $\pm{0.03 }$& 65.8 $\pm{0.06}$& 62.0 $\pm{ 0.3 }$ & 62.3 $\pm{ 0.04 }$& 66.3 $\pm{0.88}$\\
\enddata
\tablenotetext{a}{Elemental abundances relative to solar \citep{Grev96}}
\tablenotetext{b}{Note that these quantities depend on the density according to the equations 
in the text.}
\end{deluxetable}



\end{document}